\documentclass[10.5pt]{article}
\usepackage[leqno]{amsmath}
\usepackage{amsfonts}
\usepackage{graphicx}
\usepackage{caption}
\usepackage{subcaption}
\usepackage{titlesec}
\usepackage{bigints}
\usepackage{bbm}
\usepackage{amsmath}
\usepackage{amssymb}
\usepackage{latexsym}
\usepackage{blindtext}
\usepackage[linktocpage=true]{hyperref}
\usepackage{amsmath, amsfonts,amssymb, amsthm, euscript,makeidx,color,mathrsfs}
\usepackage{cite}
\usepackage{array}
\usepackage{multirow}
\usepackage{makecell}
\usepackage{pgfplots}
\usepackage{diagbox}
\usepackage{afterpage}
\usepackage[capitalize,nameinlink]{cleveref}

\usepackage{float}

\titleformat{\paragraph}
{\normalfont\normalsize\bfseries}{\theparagraph}{1em}{}
\titlespacing*{\paragraph}
{0pt}{3.25ex plus 1ex minus .2ex}{1.5ex plus .2ex}

\hypersetup{
    colorlinks=true,
    linkcolor=blue,
    citecolor=cyan,
}

\def\sqr#1#2{{\vcenter{\vbox{\hrule height.#2pt
              \hbox{\vrule width.#2pt height#1pt \kern#1pt \vrule width.#2pt}
              \hrule height.#2pt}}}}
\def\signed #1{{\unskip\nobreak\hfil\penalty50
              \hskip2em\hbox{}\nobreak\hfil#1
              \parfillskip=0pt \finalhyphendemerits=0 \par}}
\def\endpf{\signed {$\sqr69$}}

\def \Re{\mbox{\rm Re}}
\def\3n{\negthinspace \negthinspace \negthinspace }
\def\2n{\negthinspace \negthinspace }
\def\1n{\negthinspace }

\def\dbC{\mathbb{C}}

\def\dbE{\mathbb{E}}
\def\dbF{\mathbb{F}}

\def\dbN{\mathbb{N}}

\def\dbP{\mathbb{P}}

\def\dbR{\mathbb{R}}

\def\sD{\mathscr{D}}

\def\={\buildrel \triangle \over =}

\def\ds{\displaystyle}

\def\ns{\noalign{\ss}}
\def\a{\alpha}
\def\b{\beta}
\def\g{\gamma}
\def\d{\delta}
\def\e{\varepsilon}
\def\z{\zeta}
\def\k{\kappa}
\def\l{\lambda}
\def\m{\mu}
\def\n{\nu}
\def\si{\sigma}
\def\t{\tau}
\def\f{\varphi}

\def\i{\infty}
\def\G{\Gamma}
\def\D{\Delta}

\def\O{\Omega}

\def\cF{{\cal F}}

\def\BB{{\bf B}}

\def\BD{{\bf D}}

\def\BM{{\bf M}}
\def\BN{{\bf N}}

\def\BP{{\bf P}}

\def\BX{{\bf X}}

\def\BZ{{\bf Z}}

\def\BBl{\boldsymbol\lambda}
\def\BBm{\boldsymbol\mu}

\def\BBL{\boldsymbol\Lambda}

\def\G{\Gamma}
\def\D{\Delta}

\def\O{\Omega}

\def\BBL{\boldsymbol\Lambda}

\def\BBF{\boldsymbol\Phi}

\def\ss{\smallskip}
\def\ms{\medskip}

\def\q{\quad}
\def\qq{\qquad}
\def\hb{\hbox}

\def\liminf{\mathop{\underline{\rm lim}}}

\def\da{\mathop{\downarrow}}
\def\Ra{\mathop{\Rightarrow}}

\def\lan{\mathop{\langle}}
\def\ran{\mathop{\rangle}}

\def\h{\widehat}
\def\wt{\widetilde}

\def\cd{\cdot}
\def\cds{\cdots}

\def\as{\hbox{\rm a.s.{ }}}
\def\sgn{\hbox{\rm sgn$\,$}}
\def\diag{\hbox{\rm $\,$diag$\,$}}

\def\tr{\hbox{\rm tr$\,$}}

\def\les{\leqslant}
\def\ges{\geqslant}
\def\Re{{\mathop{\rm Re}\,}}

\def\({\Big (}
\def\){\Big )}
\def\[{\Big[}
\def\]{\Big]}

\def\bde{\begin{definition}\label}
	\def\ede{\end{definition}}
	\def\bel{\begin{equation}\label}
		\def\ee{\end{equation}}
	\def\bt{\begin{theorem}\label}
		\def\et{\end{theorem}}
	\def\bc{\begin{corollary}\label}
		\def\ec{\end{corollary}}
	\def\bl{\begin{lemma}\label}
		\def\el{\end{lemma}}
	\def\bp{\begin{proposition}\label}
		\def\ep{\end{proposition}}
	\def\bas{\begin{assumption}}
		\def\eas{\end{assumption}}
	\def\br{\begin{remark}\label}
		\def\er{\end{remark}}
	\def\ba{\begin{array}}
		\def\ea{\end{array}}

\def\ba{\begin{array}}
\def\ea{\end{array}}

\def\rf{\eqref}

\def\square#1{\vbox{\hrule\hbox{\vrule height#1%
     \kern#1\vrule}\hrule}}
\def\rectangle#1#2{\vbox{\hrule\hbox{\vrule height#1%
     \kern#2\vrule}\hrule}}

\font\tenbb=msbm10 \font\sevenbb=msbm7 \font\fivebb=msbm5

\newfam\bbfam
\scriptscriptfont\bbfam=\fivebb \textfont\bbfam=\tenbb
\scriptfont\bbfam=\sevenbb

\newtheorem{lemma}{Lemma}[section]
\newtheorem{remark}{Remark}[section]

\newtheorem{theorem}{Theorem}[section]
\newtheorem{corollary}{Corollary}[section]

\newtheorem{definition}{Definition}[section]
\newtheorem{proposition}{Proposition}[section]
\newtheorem{assumption}{Assumption}[section]

\makeatletter
   
   \@addtoreset{equation}{section}
\makeatother

\numberwithin{equation}{section}

\begin{document}

\title{\bf A Limit Order Book Model for High Frequency Trading with Rough Volatility}

\author{Yun Chen-Shue\thanks{Department of
Mathematics, University of Central Florida, Orlando, FL 32816, USA. Email:
{\tt yun.su.chenshue@gmail.com.}},~~~Yukun Li\thanks{Department of Mathematics, University of Central Florida, Orlando,
FL 32816, USA. Email:{\tt yukun.li@ucf.edu.} This author was partially supported by
NSF DMS-2110728.}~~~and~~~
Jiongmin Yong\thanks{Department of Mathematics, University of Central Florida, Orlando,
FL 32816, USA. Email:{\tt jiongmin.yong@ucf.edu.} This author was partially supported by
NSF DMS-2305475.}}

\maketitle

\begin{abstract}
We introduce a model for limit order book of a certain security with two main features: First, both the limit orders and market orders for the given asset are allowed to appear and interact with each other.  Second, the high frequency trading  activities are allowed and described by the scaling limit of nearly-unstable multi-dimensional Hawkes processes with power law decay.  The model has been derived as a stochastic partial differential equation (SPDE, for short), under certain intuitive identifications. Its diffusion coefficient is determined by a Volterra integral equation driven by a Hawkes process, whose Hurst exponent is less than 1/2 (so that the relevant process is negatively correlated). As a result, the volatility path of the SPDE is rougher than that driven by a (standard) Brownian motion. The well-posedness follows from a result in literature. Hence, a foundation is laid down for further studies in this direction.

\end{abstract}

\bf Keywords. \rm Limit order book, Hawkes process, rough volatility, Volterra integral equations, high frequency trading.

\ms

\bf AMS Mathematics subject classification. \rm 60H15, 91G80, 60G55, 60G22

\ms

\section{Introduction}

A limit order book (LOB, for short), a list of prices and volumes for a traded asset, can be used as a mechanism to facilitate trades in the financial market: traders can place limit orders in the order book with pre-determined prices and volumes waiting for execution as well as submit market orders that can be executed immediately against the existing limit orders by the best available prices.  For each time $t$, the LOB provides a snapshot of the market by presenting the volumes of outstanding limit orders at each price level.  The price level increments by the minimum price change is called the tick size.  In the LOB example below, the tick size is 1 cent.  The green columns visualize the volumes of the bid orders (or, buy orders) and are negative by convention.  The red columns show the volumes of the ask orders (or, sell orders) and are positive by convention also.  The highest bid offer, \$100.00 in the example, is called the bid price, while the lowest ask offer (\$100.01) is called the ask price.  The mid-price of a LOB is often calculated as the average of the bid and ask prices, which is \$100.005 in the example below.

\begin{figure}[!h]
\centering
\includegraphics[width=1\textwidth]{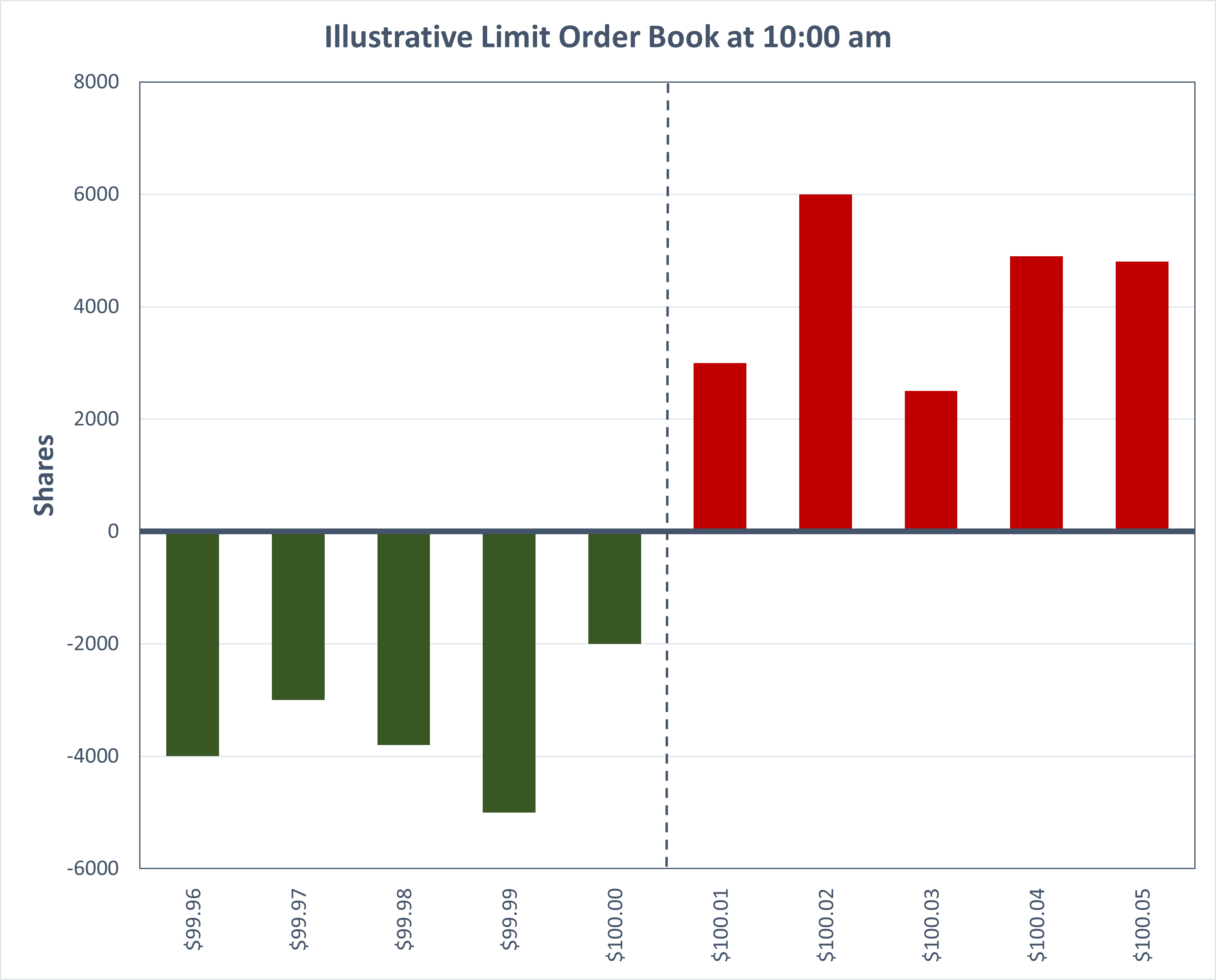}
\caption{Illustrative LOB at 10:00 am}
\label{LOB}
\end{figure}

Since the LOB dynamics shows the supply and demand of the given asset in a fundamental way and forms the price dynamics of this asset, there has been an increasing interest in modeling the LOB dynamics.  However, most modeling attempts are hard to be analytically or computationally tractable \cite{Parlour-1998}, \cite{Bouchaud-Mezard-Potters-2004}, \cite{Foucault-Kadan-Kandel-2005}, \cite{Rosu-2009}.


Cont and M\"uller \cite{Cont-Mueller-2021} proposed a model in which the dynamics of the centered order book density is described by a stochastic partial differential equation (SPDE, for short) with multiplicative Gaussian noise, which is used to described the high frequency trading (HFT, for short). We will refer to this model as the Cont-M\"uller model (C-M model, for short) in the rest of this paper.  The centered order book density, $u(t,x)$, is the volume per unit price (tick size) of the limit order at time $t$ and the position $x$ is the distance away from the mid-price, with $x \in [-L,L]$ for some $L>0$.  It is easy to see that rational investors will not submit limit orders far away from the mid-price,
and that all the previously-submitted orders were cancelled as soon as their price levels became too far away from the mid-price.  This assumption is reflected by  setting $u(t,x)=0$ when $x\notin(-L,L)$ (See \cite{Cont-Mueller-2021}).

\ms

The C-M model that presents the centered volume density $u(t,x)$ can be written as follows (see \cite{Cont-Mueller-2021}, with small modifications):
$$\ba{ll}
\ns\ds du(t,x)=\big[\eta_a\D u(t,x)+\b_a\nabla u(t,x)-\a_a u(t,x)+f^a(x)\big]dt+\si_a u(t,x)dW^a(t),\qq x\in(0,L),\\
\ns\ds du(t,x)=\big[\eta_b\D u(t,x)-\b_b\nabla u(t,x)-\a_b u(t,x)-f^b(x)\big]dt+\si_b u(t,x)dW^b(t),\qq x\in(-L,0),\ea$$
with
$$\ba{ll}
\ns\ds u(t,x)\les0,\q x<0,\qq u(t,x)\ges0,\q x>0,\\
\ns\ds u(t,0+)=u(t,0-)=0,\qq u(t,-L)=u(t,L)=0,\ea$$
where $\eta_a,\eta_b,\b_a,\b_b,\si_a,\si_b,\a_a,\a_b>0$ are some constants, $f^a,f^b: [-L,L]\to[0,\i)$ are given functions, and $(W^a,W^b)$ is a two-dimensional Brownian motion (with possibly correlated components). In these equations, non-high frequency trading (non-HFT, for short) order submissions are modeled by $f^a(x)$ and $f^b(x)$, all kinds of non-HFT order cancellations/replacements by
$$\big[\eta_a\D u(t,x)+\b_a\nabla u(t,x)-\a_a u(t,x)\big],\qq\big[\eta_b\D u(t,x)-\b_b \nabla u(t,x)-\a_b u(t,x)\big],$$
and HFT order dynamics by $\si_a u(t,x)dW^a(t)$ and $\si_bu(t,x)dW^b(t)$, on the ask and bid sides, respectively.  We will provide detailed explanations of the relevant terms when introducing our model in Section 3.

\ms

The C-M model \cite{Cont-Mueller-2021} has both the analytical and computational tractability for applications, and the price dynamics was naturally derived from the model.  However, there are two main limitations in that model.

\ms

First, the C-M model did not reflect the effect to the centered order book density from market ask/bid orders. Indeed, the only terms regarding order submissions are $f^a(x)$ and $f^b(x)$, which only increase the volumes on the ask and bid sides, whereas the market order submissions affect the LOB in a different way since they decrease the LOB volumes. Thus, the market orders should be taken into account.
\ms

Second, the C-M model used multiplicative Gaussian noise terms to model the order dynamics from HFT at coarse-grained time scale of the average (non-HFT) market participants. This implies that each increment of the HFT is independent of the previous HFT incremental changes. However, many evidence shows that HFT markets are highly endogenous, meaning HFT orders tend to generate other HFT orders. Furthermore, many HFT orders are part of a larger order (or metaorder) that takes a relatively long time to fully execute, which causes a given HFT order to have a relatively long-term influence on other HFT orders.  Thus, it is better to use self-exciting and long term dependency process to model HFT, rather than Brownian motions (as in the C-M model) \cite{El_Euch-Fukasawa-Rosenbaum-2018}.

\ms

In this paper, we propose a new model. First, we include the effect from market orders so that the limit orders and market orders interact with each other, which looks more realistic. Second, we have used the scaling limit of a nearly-unstable multivariate Hawkes process sequence (which is self-exciting among all components) with power-law tails to model the HFT dynamics at a coarse-grained time scale, reflecting the dependencies among HFT orders.

\ms

The remaining of this paper is arranged as follows. Section 2 provides a brief overview on the Hawkes process. Section 3 presents our new model with both the non-HFT and HFT components. Section 4 gives a derivation for the scaling limit of some processes relevant to the Hawkes processes, while Section 5 presents the SPDE of the market model and its well-posedness. In Section 6, we derive the price dynamics based on the order book dynamics. Some numerical result are collected in Section 7. Finally, a technical result will be put in the appendix.

\ms

For relevant results,the readers are referred to \cite{Parlour-1998}, \cite{Bouchaud-Mezard-Potters-2004}, \cite{Rosu-2009}, \cite{Cont-Larrard-2013}, \cite{Cont-Kukanov-Stoikov-2014}, \cite{Said-Ayed-Husson-Abergel-2018}, \cite{Cont-Mueller-2021} for modeling LOB, to \cite{Bacry-Dayri-Muzy-2012}, \cite{Menkveld-2013}, \cite{Kirilenko-Kyle-Samadi-Tuzun-2014}, \cite{Bacry-Jaisson-Muzy-2016}, \cite{Subrahmanyam-Zheng-2016},
\cite{Lu-Abergel-2018}, \cite{Said-Ayed-Husson-Abergel-2018}, \cite{Manahov-2021}, \cite{Protter-Wu-Yang-2021}
for modeling HFT, to \cite{Hawkes-1971}, \cite{Hawkes-Oakes-1974}, \cite{Bremaud-Massoulie-1996}, \cite{Caestensen-2010}, \cite{Bacry-Delattre-Hoffmann-Muzy-2013a},
\cite{Bacry-Delattre-Hoffmann-Muzy-2013b}, \cite{Hardiman-Bercot-Bouchaud-2014}, \cite{Laub-Taimre-Pollet-2015}, \cite{Jaisson-Rosenbaum-2015}, \cite{Jaisson-Rosenbaum-2016}, \cite{Kirchner-2016}, \cite{Laub-Lee-Taimre-2021} for Hawkes process related results, to \cite{Heston-1993}, \cite{Gatheral-Jaisson-Rosenbaum-2018} for rough volatility.

\section{An Overview of the Hawkes Process}

In this section, we briefly provide an overview of the $m$-dimensional Hawkes process (with $m\ges1$). Most of the following definitions and propositions are from \cite{Laub-Lee-Taimre-2021}.

\subsection{One-dimensional Hawkes process}

\bde{exponential} \rm (i) A discrete random variable $X$ is said to have a {\it Poisson distribution} with a parameter $\l^*>0$ if it has a discrete probability distribution:
$$f(k;\l^*)=\dbP(X=k)={(\l^*)^ke^{-\l^*}\over k!}, \qq\forall k=0,1,2,...$$
We denote it as $X\sim{\rm Poi}(\l^*)$.

\ms

(ii) A {\it counting process} is a stochastic process $(N(t):t\ges0)$ taking values in the set $\{0,1,2,...\}$ that satisfies $N(0)=0$, almost surely finite, and is a right-continuous non-decreasing step function with increments of size $+1$.
Further, denote by $\dbF=\{\cF_t\}_{t\ges0}$ a right-continuous {\it filtration}, that is, an increasing sequence of $\si$-field, such that $\cF_t=\bigcap_{\e>0}\cF_{t+\e}$. The filtration $\dbF$ represents the history of the counting process $N(\cd)$, namely, it is generated by $N(\cd)$.

\ms

(iii) A counting process $N(\cd)$ is called a (an {\it inhomogeneous}) {\it Poisson process} with {\it intensity function} (or {\it rate function}) $\l(t)>0$ if it has independent increments and for any interval $I=(a,b]$, $N(I)\equiv\{N(t)\bigm|t\in(a,b]\}$ has a Poisson distribution with parameter $\int^b_a\l(s)ds$, i.e.,
$$N(I)\sim{\rm Poi}\Big(\int^b_a\l(s)ds\Big),\q\hb{or}\q\dbP\big(N(I)=k\big)= {\ds\(\int^b_a\l(s)ds\)^ke^{-(\int^b_a \l(s)ds)}\over k!},\qq\forall k=0,1,2,...$$
If the intensity function is a constant $\l>0$, then $N(\cd)$ is called a {\it homogeneous Poisson process}.

\ms

(iv) Let $N(\cd)$ be a counting process whose histories are described by $\dbF=\{\cF_t\}_{t\ges0}$. If a (non-negative) function $\l(t)$ exists such that
\bel{conditional intensity}\l(t)=\lim_{h\da0}{\dbE\big[N(t+h)-N(t)\,|\,\cF_t\big]\over h},\qq t\ges0,\ee
is well-defined. It is called the {\it conditional intensity function} of $N(\cd)$.
Consequently,
\bel{E[dl]}\dbE[dN(t)]=\dbE[\l(t)]dt.\ee

\ede

\bde{Hawkes} \rm A counting process $(N(t):t\ges0)$ is called a {\it Hawkes process} if the following conditions hold:

\ms

(a) The conditional increment against its history $\dbF=\{\cF_t\}_{t\ges0}$ satisfies
$$\dbP\big(N(t+h)-N(t)=k\,|\,\cF_t\big)=\left\{\ba{ll}
\ns\ds1-\l(t)h+o(h),\qq\;k=0,\\
\ns\ds\l(t)h+o(h),\qq\qq k=1,\\
\ns\ds o(h),\qq\qq\qq\q\;\; k>1,\ea\right.$$
for some conditional intensity function $\l(\cd)$, which is non-negative valued.

\ms

(b) The conditional intensity function $\l(\cd)$ is of the form
\bel{l=}\l(t)=\m(t)+\int^t_0\phi(t-s)dN(s),\qq t\ges0,\ee
where $\m(t)$, called the {\it background intensity}, or the intensity function of an underlying Poisson process, is a deterministic positive bounded function of $t$ that has a finite limit $\m(\i)>0$ as $t\to\i$, and $\phi:(0,\i)\to[0,\i)$, called the {\it excitation function}, is assumed to be a deterministic positive bounded function. Because of the above, $\l(\cd)$ is positively valued.

\ede

The exogenous events arrival is described by an inhomogeneous Poisson process with the rate function $\m(t)>0$, and the direct offspring of any event arrival is described by an inhomogeneous Poisson process with the rate function being the integral of function $\phi(\cd)$ with respect to the Hawkes process $N(\cd)$ itself, showing the feature of self-exciting. Intuitively, $\phi(\cd)$ measures the effect of previous arriving events on the latter ones. Thus, it is natural to assume that such a function to be non-negative (and non-increasing because of the fading memory). However, in the sequel, we do not need such a monotonicity assumption.

\ms

Note that if $\phi(\cd)=0$, the Hawkes process $N(\cd)$ becomes an inhomogeneous Poisson
process.  Thus the former can be regarded an extension of the latter mathematically.
According to \cite{Hawkes-Oakes-1974}, \cite{Westcott-1977}, and \cite{Kirchner-2016},
the above-defined Hawkes process exists as long as the following {\it stability condition}
is satisfied:
\bel{|f|<1}\|\phi(\cd)\|_1\equiv\int^\i_0\phi(t)dt<1.\ee

By definition, any given Hawkes process $N(\cd)$ is c\`adl\`ag and non-decreasing. Moreover its intensity
\bel{l>0}\l(t)>0,\qq t\ges0.\ee
These facts will be used later. Next, we let
\bel{M}M(t)=N(t)-\int_0^t\l(s)ds,\qq t\ges0.\ee
Then it is a martingale and the following holds
\bel{MM}\lan M,M\ran(t)=\int_0^t\l(s)ds,\qq t\ges0.\ee
Namely, $t\mapsto M(t)^2-\lan M,M\ran(t)$ is a martingale (by Doob-Meyer decomposition theorem). In fact, note that the jump locations and sizes of $M(\cd)$ and $N(\cd)$ are the same:
$$\D M(t)=\D N(t)=0,1.$$
Thus,
$$\D M(t)^2=\D N(t)^2=\D N(t).$$
Hence, by \cite{Protter-2005}, p.78, Theorem 31, one has
$$\ba{ll}
\ns\ds M(t)^2=\int_{0+}^t2M(s-)dM(s)+\sum_{0<s\les t}[M(s)^2-M(s-)^2-2M(s-)\D M(s)]\\
\ns\ds\qq=\int_{0+}^t2M(s-)dM(s)+\sum_{0<s\les t}\D M(s)^2=\int_{0+}^t2M(s-)dM(s)+\sum_{0<s\les t}\D N(s)\\
\ns\ds\qq=\int_{0+}^t2M(s-)dM(s)+N(t),\qq t\ges0.\ea$$
Consequently,
\bel{EM^2=EN}\dbE[M(t)^2]=\dbE[N(t)]=\int_0^t\dbE[\l(s)]ds,\qq t\ges0.\ee
That implies $M(t)^2-\int_0^t\l(s)ds$ is a martingale, proving \rf{MM}. Then it is useful that
\bel{E|psidM|}\dbE\Big|\int_0^t\psi(s)dM(s)\Big|^2=\int_0^t|\psi(s)|^2\l(s)ds,\ee
provided both sides make sense.

\ms

By the definition of $M(\cd)$, \rf{l=} can be written as
\bel{l=*}\l(t)=\m(t)+\int^t_0\phi(t-s)\l(s)ds+\int_0^t\phi(t-s)dM(s),\qq t\ges0,\ee
Consequently, we have (noting $\m(\cd)$ and $\phi(\cd)$ are deterministic)
\bel{El=}\dbE\big[\l(t)\big]=\m(t)+\int^t_0\phi(t-s)\dbE\big[\l(s)\big]ds=\m(t)
+\int_0^t\phi(s)\dbE\big[\l(t-s)\big]ds,\qq t\ges0.\ee
This is a Volterra integral equation which is always globally solvable, under the boundedness of $\phi(\cd)$. Moreover, the solution is positive since both $\m(t)$ and $\phi(t)$ are positive. This also follows from \rf{l>0}. By Gronwall inequality, one has
\bel{El<}0\les\dbE\big[\l(t)\big]\les\|\m(\cd)\|_\i e^{\|\phi(\cd)\|_\i t},\qq t\ges0.\ee
Thus, $\dbE\big[\l(t)\big]$ is bounded in any finite interval, at least. Now, the limit $\ds\lim_{t\to\i}\dbE\big[\l(t)\big]$ might exist or not. In the former case, we denote
\bel{liml(t)}\lim_{t\to\i}\dbE\big[\l(t)\big]=\dbE\big[\l(\i)\big]>0.\ee
This is necessary for $\l(t)$ to be asymptotically stationary. When \rf{liml(t)} holds, we say that the Hawkes process $N(\cd)$ is {\it stable}. Otherwise, we say that the Hawkes process $N(\cd)$ (it might not exist, and even it exists) is {\it unstable}. In the case that \rf{liml(t)} holds, by the dominated convergence theorem, we have (from \rf{El=})
$$\dbE\big[\l(\i)\big]=\m(\i)+\|\phi(\cd)\|_1\dbE\big[\l(\i)\big].$$
Since $\m(\i)>0$, from the above, we see that \rf{|f|<1} must be true, and
\bel{El(i)=}\dbE\big[\l(\i)\big]={\m(\i)\over1-\|\phi(\cd)\|_1}.\ee
Thus, if \rf{|f|<1} is not satisfied, the Hawkes process, even if it exists, is unstable.
In the above, we a priori assume the existence of the limit in \rf{liml(t)} to obtain \rf{|f|<1} and \rf{El(i)=}. We now present a result, which is another way around.

\bp{hawkes-process-long-term-expectation} \sl Let $N(\cd)$ be a one-dimensional Hawkes process whose conditional intensity process has the form \rf{l=} with $\m(\cd)$ and $\phi(\cd)$ given as in Definition \ref{Hawkes}. Let the stability condition \rf{|f|<1} hold. Then,
\bel{|bar l|}\big\|\dbE\big[\l(\cd)\big]\big\|_\i\equiv\sup_{t\ges0}\big|\dbE\big[\l(t)\big]\big| \les{\|\m(\cd)\|_\i\over1-\|\phi(\cd)\|_1},\ee
which is an improvement of \rf{El<}. Moreover, \rf{liml(t)} and \rf{El(i)=} also hold. Consequently,
\bel{EN(t)}\dbE[N(t)]\les{\|\m(\cd)\|_\i\over1-\|\phi(\cd)\|_1}t,\qq t\ges0.\ee
Further, without assuming the stability condition \rf{|f|<1}, for any $T>0$ with $\|\phi(\cd){\bf1}_{[0,T]}\|_1<1$, it holds
\bel{bar l>}\liminf_{t\to\i}
\dbE\big[\l(t)\big]\ges{\m(\i)\over1-\|\phi(\cd){\bf1}_{[0,T]}\|_1}\(1+{
\ds\int_T^\i\phi(s)ds\over1-\|\phi(\cd){\bf1}_{[0,T]}\|_1}\).\ee
Consequently, in the case that the stability condition \rf{|f|<1} fails, although $\dbE\big[\l(\cd)\big]$ is still well-defined as the solution of Volterra integral equation \rf{El=} over $[0,\i)$, it holds
$$\lim_{t\to\i}\dbE\big[\l(t)\big]=\i.$$

\ep

\it Proof. \rm Denote $\bar\l(t)=\dbE\big[\l(t)\big]$. Then \rf{El=} can be written as
\bel{bar l}\bar\l(t)=\m(t)+\int^t_0\phi(t-s)\bar\l(s)ds=\m(t)+[\phi*\bar\l](t),\qq t\ges0.\ee
By the boundedness of $\m(\cd)$ and the Young's inequality, one has
$$\|\bar\l(\cd)\|_\i\les\|\m(\cd)\|_\i+\|\phi(\cd)\|_1
\|\bar\l(\cd)\|_\i.$$
Thus, by the stability condition \rf{|f|<1}, we have \rf{|bar l|}. Let us further refine it. Note that
\bel{bar l}\ba{ll}
\ns\ds\bar\l(t)=\big[\m+\phi*\bar\l\big](t)=\big[\m+\phi*(\m+\phi*\bar\l)\big](t)=\cds
=\sum_{k=0}^m\big[\phi^{*k}*\m\big](t)+\big[\phi^{*(m+1)}*\bar\l\big](t),\ea\ee
where (note $\phi(\cd)$ is non-negative)
$$\ba{ll}
\ns\ds\phi^{*0}(t)=1,\q\phi^{*1}(t)=\phi(t),\qq t\ges0,\\
\ns\ds\phi^{*k}(t)=[\phi^{*(k-1)}*\phi](t)=\int_0^t\phi^{*(k-1)}(t-s)\phi(s)ds\ges0,\q t\ges0,\q k\ges2.\ea$$
In what follows, we will always naturally extend 0 on $(-\i,0)$ for functions defined on $[0,\i)$. Thus, $\phi^{*k}(t)=0$, for $t<0$ and $k\ges1$. Note that
\bel{int f}\ba{ll}
\ns\ds0\les\int_0^t\phi^{*k}(r)dr=\int_0^t\[\int_0^r\phi^{*(k-1)}(r-s)\phi(s)ds\]dr\\
\ns\ds\q=\int_0^t\int_s^t\phi^{*(k-1)}(r-s)\phi(s)drds
=\int_0^t\(\int_0^{t-s}\phi^{*(k-1)}(r)dr\)\phi(s)ds\\
\ns\ds\q\les\int_0^t\(\int_0^t\phi^{*(k-1)}(r)dr\)\phi(s)ds\les\cds\\
\ns\ds\q\les\(\int_0^t\phi(s)ds\)^{k-1}\(\int_0^t\phi(t)dt\)=\(\int_0^t\phi(t)dt\)^k,\ea\ee
and
\bel{|f|_1}\ba{ll}
\ns\ds\|\phi^{*k}(\cd)\|_1=\int_0^\i\phi^{*k}(t)dt=\int_0^\i\[\int_0^t\phi^{*(k-1)}(t-s)\phi(s)ds\]dt\\
\ns\ds=\int_0^\i\int_s^\i\phi^{*(k-1)}(t-s)\phi(s)dtds
=\int_0^\i\phi^{*(k-1)}(t)dt\int_0^\i\phi(s)ds=\cds\\
\ns\ds=\(\int_0^\i\phi(s)ds\)^{k-1}\(\int_0^\i\phi(t)dt\)=\(\int_0^\i\phi(t)dt\)^k
=\|\phi(\cd)\|_1^k.\ea\ee
Thus, by Young's inequality again, we have
\bel{f^km}\|[\phi^{*k}*\m](\cd)\|_\i\les\|\phi^{*k}(\cd)\|_1\|\m(\cd)\|_\i=\|\phi(\cd)\|_1^k
\|\m(\cd)\|_\i,\ee
and (noting \rf{|bar l|})
\bel{f^kl}\|[\phi^{*(m+1)}*\bar\l](\cd)\|_\i\les\|\phi^{*(m+1)}(\cd)\|_1\|\bar\l(\cd)\|_\i\les
\|\phi(\cd)\|_1^{m+1}{\|\m(\cd)\|_\i\over1-\|\phi(\cd)\|_1}.\ee
Hence, as $m\to\i$, the series in \rf{bar l} is convergent, and the last term goes to zero. Consequently,
\bel{bar l*}\bar\l(t)=\sum_{k=0}^\i\big[\phi^{*k}*\m\big](t),\qq t\ges0.\ee
Next, let $t\to\i$, noting $\m(t)\to\m(\i)$, by the dominated convergence and monotone convergence theorems, we have
$$\lim_{t\to\i}\dbE[\l(t)]=\sum_{k=0}^\i\lim_{t\to\i}\int^t_0\m(t-s)\phi^{*k}(s)ds  =\m(\i)\sum_{k=0}^\i\|\phi^{*k}(\cd)\|_1=\m(\i)\sum_{k=0}^\i\|\phi(\cd)\|_1^k={\m(\i)
\over1-\|\phi(\cd)\|_1}.$$
This proves our conclusion for the case that the stability condition \rf{|f|<1} is assumed. Relation \rf{EN(t)} is clear.

\ms

Now, without assuming \rf{|f|<1}, for $T>0$ with $\|\phi(\cd){\bf1}_{[0,T]}\|_1<1$, we set
$$\phi_T(t)=\phi(t){\bf1}_{[0,T]}(t),\qq t\ges0,$$
and define
$$\bar\l_T(t)=\m(t)+\int_0^t\phi_T(t-s)\bar\l(s)ds=\sum_{k=0}^\i[\phi_T^{*k}*\m](t),\qq t\ges1.$$
Note that since $\m(\cd)>0$, we have
$$\int_0^t\phi_T(s)\m(t-s)ds\les\int_0^t\phi(s)\m(t-s)ds,\qq t\ges0.$$
Hence, by induction and \rf{bar l*}, one sees that
$$\bar\l_T(t)\les\bar\l(t),\qq\forall T,t\ges0.$$
Applying the above proved conclusion with $\phi(\cd)$ replaced by $\phi(\cd){\bf1}_{[0,T]}(\cd)$, we have
$$\lim_{t\to\i}\bar\l_T(t)={\m(\i)\over1-\|\phi_T(\cd)\|_1}.$$
On the other hand,
$$\bar\l(t)-\bar\l_T(t)=\int_0^t\(\phi(s)-\phi_T(s)\)\bar\l(t-s)ds+\int_0^t\phi_T(s)
\big[\bar\l(t-s)-\bar\l_T(t-s)\big]ds,$$
This implies
$$\lim_{t\to\i}\big[\bar\l(t)-\bar\l_T(t)\big]={\ds\liminf_{t\to\i}\int_T^t\phi(s)\bar\l(t-s)ds
\over1-\|\phi_T\|_1}\ges{\ds\liminf_{t\to\i}\int_T^t\phi(s)\bar\l_T(t-s)ds
\over1-\|\phi_T\|_1}={
\ds\m(\i)\int_T^\i\phi(s)ds\over(1-\|\phi_T\|_1)^2}.$$
Hence,
$$\liminf_{t\to\i}\bar\l(t)\ges{\m(\i)\over1-\|\phi_T(\cd)\|_1}\(1+{
\ds\int_T^\i\phi(s)ds\over1-\|\phi_T\|_1}\).$$
Therefore, \rf{bar l>} follows. \endpf

\ms

From \rf{bar l>}, we see that if the condition
\bel{|f|=1}\|\phi(\cd)\|_1=1\ee
holds, and $\phi(\cd)$ is strictly positive, then for any $T>0$, $\|\phi(\cd){\bf1}_{[0,T]}\|_1<1$. Thus, \rf{bar l>} holds for any $T>0$, which means $\dbE[\l(t)]$ is (still) well-defined for each $t>0$, but it blows up when $t\to\i$. We will thus call \rf{|f|=1} the {\it critical unstable condition} from now on.
Further, we have the following result.

\bp{} \sl Let the stability condition \rf{|f|<1} hold. Then
\bel{l(t)=}\l(t)=\sum_{k=0}^\i[\phi^{*k}*\m](t)+\sum_{k=1}^\i[\phi^{*k}*dM](t),\qq t\ges0.\ee
Moreover, for any $\psi(\cd)\in L^2(0,\i)$,
\bel{psi*dM}\dbE|[\psi*dM](t)|\les{\|\m(\cd)\|_\i^{1\over2}
\|\psi(\cd)\|_2\over(1-\|\phi(\cd)\|_1)^{1\over2}},\ee
where $\ds\|\psi(\cd)\|_2=\(\int_0^\i|\psi(s)|^2ds\)^{1\over2}$. In particular,
\bel{f*dM}\dbE|[\phi^{*k}*dM](t)|\les\({\|\m(\cd)\|_\i\over1-\|\phi(\cd)\|_1}\)^{1\over2}
\|\phi(\cd)\|_2\|\phi(\cd)\|_1^{k-1},\q k\ges1.\ee

\ep

\it Proof. \rm By induction, we have from \rf{l=*} that
\bel{l(t)=*}\ba{ll}
\ns\ds\l(t)=\m(t)+\int^t_0\phi(t-s)\l(s)ds+\int_0^t\phi(t-s)dM(s)\\
\ns\ds\qq=\m(t)\1n+\2n\int^t_0\3n\phi(t-s)\big[\m(s)\1n+\2n\int^s_0\3n\phi(s-r)\l(r)dr
+\2n\int_0^s\3n\phi(s-r)
dM(r)\big]dr+\2n\int_0^t\3n\phi(t-s)dM(s)=\cds\\
\ns\ds\qq=\sum_{k=0}^m[\phi^{*k}*\m](t)+\sum_{k=1}^{m+1}[\phi^{*k}*dM](t)
+[\phi^{*(m+1)}*\l](t).\ea\ee
By \rf{f^km}, the first series is convergent. Similar to \rf{f^kl}, the last term goes to zero almost surely. By \rf{|bar l|} and Young's inequality, we have
$$\ba{ll}
\ns\ds\dbE|[\psi*dM](t)|=\dbE\Big|\int_0^t\psi(t-s)dM(s)\Big|=\(\dbE\Big|\int_0^t
\psi(t-s)dM(s)\Big|^2\)^{1\over2}\\
\ns\ds\qq\qq\qq\q=\(\dbE\int_0^t|\psi(t-s)|^2\l(s)ds\)^{1\over2}\les
\|\dbE[\l(\cd)]\|_\i^{1\over2}\|\psi(\cd)^2\|_1^{1\over2}\les\({\|\m(\cd)\|_\i\over1-\|\phi(\cd)\|_1}\)^{1\over2}
\|\psi(\cd)\|_2.\ea$$
This proves \rf{psi*dM}. In particular,
$$\dbE|[\phi^{*k}*dM](t)|\les\({\|\m(\cd)\|_\i\over1-\|\phi(\cd)\|_1}\)^{1\over2}
\|\phi^{*k}(\cd)\|_2\les\({\|\m(\cd)\|_\i\over1-\|\phi(\cd)\|_1}\)^{1\over2}
\|\phi(\cd)\|_2\|\phi(\cd)\|_1^{k-1},\q k\ges1.$$
Hence, the second series on the right-hand side of \rf{l(t)=*} is convergent as well. Therefore, \rf{l(t)=} and \rf{f*dM} are proved. \endpf

\subsection{Multi-dimensional Hawkes process}

An extension of the one-dimensional Hawkes process is a multi-dimensional space valued Hawkes process.  In \cite{Laub-Lee-Taimre-2021}, the term multi-dimensional Hawkes process was reserved only for the multi-dimensional space valued process where the components are decoupled, and hence the components are not mutually exciting.  Meanwhile, in \cite{Andersen-2014},  \cite{Bacry-Mastromatteo-Muzy-2015}, \cite{Bacry-Jaisson-Muzy-2016}, \cite{Lu-Abergel-2018}, \cite{El_Euch-Fukasawa-Rosenbaum-2018}, it was assumed that the components are coupled so that they are mutually exciting.  In our paper, by multi-dimensional Hawkes process, we mean the process is not only multidimensional but also mutually exciting.  More precisely, we have the following definition.

\bde{def.2.6} \rm A vector-valued counting process $\BN(\cd)\equiv\big(\BN^1(\cd),..., \BN^m(\cd)\big)$ is called a {\it multi-dimensional Hawkes process}, if for each $i=1,...,m$, $\BN^i(t)$ has a conditional intensity of the form
$$\BBl^i(t)=\BBm^i(t)+\sum_{j=1}^m\int^t_0\phi^{ij}(t-s)d\BN^j(s),$$
for some positive function $\BBm^i(t)$ with $\ds\lim_{t\to\i}\BBm^i(t)=\BBm^i(\i)>0$, $\phi^{ij}:(0,\i)\to[0,\i)$, and $\phi^{ij}(\cd)\in L^1(0,\i)\cap L^\i(0,\i)$.

\ede

We can also write conditional intensity of the multi-dimensional Hawkes process in vector form as
\bel{BBl}\BBl(t)=\BBm(t)+\int_0^t\BBF(t-s)d\BN(s),\qq t\ges0.\ee
Note that $\BBm(\cd)$ is an $m$-dimensional vector-valued (deterministic, with each component being positive) function, $\BBl(\cd)$ is an $m$-dimensional (stochastic) processes with each component being positive, and $\BBF(\cd)$ is an $m\times m$ square matrix-valued (deterministic) function with non-negative entries $\phi^{ij}(\cd)$. In what follows, we are not going to study the most general case, instead, we will only consider the following special case of $m=4$ and:
\bel{F}\BBF(\cd)=\f(\cd)\BBF_0,\qq\f(\cd)\ges0,\qq\|\f(\cd)\|_1<\i.\ee
where $\BBF_0$ (and $\BBF_0^\top$) has four distinct eigenvalues $\l_1>\l_2>\l_3>\l_4>0$. Thus,
$\BBF_0$ (and $\BBF_0^\top$) is diagonalizable. Consequently, if we let $v_i$ be an eigenvector of $\BBF_0^\top$ corresponding $\l_i$ and let
\bel{P}\BP=(v_1\;v_2\;v_3\;v_4)\in\dbR^{4\times4},\ee
then $\BP$ is invertible, and
\bel{PFP}\BP^{-1}\BBF_0^\top\BP=\BD\equiv\begin{pmatrix}
        \l_1 & 0 & 0 & 0 \\
        0 & \l_2 & 0 & 0 \\
        0 & 0 & \l_3 & 0 \\
        0 & 0 & 0 & \l_4
        \end{pmatrix}.\ee
In this case, the spectrum radius $\rho\big(\BBF(t)\big)$ of the matrix function $\BBF(t)$ for each $t\ges0$ is given by
$$\rho\big(\BBF(t)\big)=\l_1\f(t),\qq t\ges0.$$
Like the one-dimensional case, if we define
\bel{BM}\BM(t)=\BN(t)-\int_0^t\BBl(s)ds,\qq t\ges0,\ee
it is a martingale, and we assume that (see
\cite{El_Euch-Fukasawa-Rosenbaum-2018}, p.253, and \cite{El_Euch-2018}, p.44)
\bel{BMBM}\lan\BM,\BM\ran(t)=\int_0^t\diag\BBl(s)ds,\qq t\ges0,\ee
so that $\BM(t)\BM(t)^\top-\lan\BM,\BM\ran(t)$ is a (square symmetric matrix
valued) martingale, following the Doob-Meyer decomposition theorem. Thus,
\bel{}\dbE\lan\BM,\BM\ran(t)=\dbE\int_0^t\diag\BBl(s)ds=\dbE\[\diag\BN(t)\].\ee
In particular,
\bel{M^iM^i}\lan\BM^i,\BM^i\ran(t)=\int_0^t\BBl^i(s)ds,\qq t\ges0,\q1\les i\les4.\ee
and
\bel{EM^iM^i}\dbE\lan\BM^i,\BM^i\ran(t)=\int_0^t\dbE\BBl^i(s)ds=\dbE\BN^i(t),\qq t\ges0,\q1\les i\les4.\ee
Also,
\bel{EM^2}\dbE\Big|\int_0^t\psi(s)^\top d\BM(s)\Big|^2=\dbE\int_0^t|\psi(s)|^2\mathbbm{1}^\top\BBl(s)ds,\qq t\ges0,\ee
provided both sides make sense. Now, \rf{BBl} can also be written as
\bel{BBl*}\BBl(t)=\BBm(t)+\int_0^t\BBF(t-s)\BBl(s)ds+\int_0^t\BBF(t-s)d\BM(s),\qq t\ges0.\ee
Note that \rf{BMBM} implies that the components of $\BM(\cd)$ are mutually independent.
However, as long as $\BBF(\cd)$ is not a diagonal matrix, the components of the intensity
$\BBl(\cd)$ are coupled, not independent, which implies that the process is self-exciting among all the components. Similar to one-dimensional case, we have the following result. The proof is parallel to that in the last subsection.

\bp{} \sl For the above Hawkes process $\BN(\cd)$, the following are true:

\ms

{\rm(i)} If for $i=1,2,3,4$, the following condition holds
\bel{l_i|f|<1}\l_i\|\f(\cd)\|_1<1,\ee
then
\bel{v_iBBl(t)=}v_i^\top\BBl(t)=\sum_{k=0}^\i\l_i^k[\f^{*k}*v_i^\top\BBm](t)+\sum_{k=1}^\i\l_i^k
[\f^{*k}*d(v_i^\top\BM)](t),\qq t\ges0,\ee
and
\bel{v_iBN(t)=}v_i^\top\BN(t)=\sum_{k=0}^\i\l_i^k\int_0^t[\f^{*k}*v_i^\top\BBm](r)dr
+\sum_{k=0}^\i\l_i^k\int_0^t[\f^{*k}*d(v_i^\top\BM)](r)dr,\qq t\ges0,\ee
with the involved series absolutely convergent. Moreover,
\bel{Ev_iBBl}\|\dbE[v_i^\top\BBl(\cd)]\|_\i\les{\|v_i^\top\BBm(\cd)\|_\i\over1-\l_i\|\f(\cd)\|_1},\ee
and
\bel{Ev_iBN}|\dbE[v_i^\top\BN(t)]|\les{\|v_i^\top\BBm(\cd)\|_\i t\over1-\l_i\|\f(\cd)\|_1},\qq t\ges0.\ee
Further, for any $\psi(\cd)\in L^2(0,\i)$,
\bel{E|psi*dBM|}\dbE\big|[\psi*d(v_i^\top\BM)](t)\big|\les\({\|v_i^\top\BBm(\cd)\|_\i\over1-\l_i\|\f(\cd)
\|_1}\)^{1\over2}\|\psi(\cd)\|_2,\ee
and
\bel{E|f*dBM|}\dbE\big|[\f^{*k}*d(v_i^\top\BM)](t)\big|\les\({\|v_i^\top\BBm(\cd)\|_\i\over1-\l_i\|\f(\cd)
\|_1}\)^{1\over2}\|\f(\cd)\|_2\|\f(\cd)\|_1^{k-1}.\ee
In particular, if \rf{l_i|f|<1} holds for $i=1$, then
\bel{BBl(t)=}\BBl(t)=\sum_{k=0}^\i\BBF_0^k[\f^{*k}*\BBm](t)+\sum_{k=1}^\i
\BBF_0^k[\f^{*k}*d\BM](t),\qq t\ges0,\ee
and
\bel{BN(t)=}\BN(t)=\sum_{k=0}^\i\BBF_0^k\int_0^t[\f^{*k}*\BBm](r)dr
+\sum_{k=0}^\i\BBF_0^k\int_0^t[\f^{*k}*d\BM](r)dr,\qq t\ges0,\ee
Consequently,
\bel{EBBl}\|\dbE[\BBl(\cd)]\|_\i\les{C\|\BBm(\cd)\|_\i\over1-\l_1\|\f(\cd)\|_1},\ee
and
\bel{EBN}|\dbE[\BN(t)]|\les{C\|\BBm(\cd)\|_\i t\over1-\l_1\|\f(\cd)\|_1},\qq t\ges0,\ee
for some absolute constant $C>0$. Hereafter, such a constant $C$ could be different from line to line.

\ms

{\rm(ii)} Let the following critical unstable condition holds
\bel{l_1|f|=1}\|\rho\big(\BBF(\cd)\big)\|_1\equiv\l_1\|\f(\cd)\|_1=1.\qq\qq(\hb{Thus, $\l_i\|\f(\cd)\|_1<1$, $i=2,3,4$.})\ee
Then
\bel{E[i(i)]*}\lim_{t\to\i}v_i^\top\BBl(t)={v_i^\top\BBm(\i)\over1-\l_i\|\f(\cd)\|_1},\qq i=2,3,4,\ee
and for any $T>0$ with $\l_1\|\f(\cd){\bf1}_{[0,T]}\|_1<1$, it holds
\bel{bar l>*}\liminf_{t\to\i}
\dbE[v_1^\top\BBl(t)]\ges{v_1^\top\BBm(\i)\over1-\l_1\|\f(\cd){\bf1}_{[0,T]}\|_1}\(1+{
\ds\l_1\int_T^\i\phi(s)ds\over1-\l_1\|\f(\cd){\bf1}_{[0,T]}\|_1}\).\ee

\ep

\br{} \rm In our definition of the multi-dimensional Hawkes process, the background intensity $\boldsymbol{\mu}(t)$ is allowed to be a vector-valued function that converges to a constant vector with positive components as $t \rightarrow \infty$.
This is slightly more general than that of \cite{Laub-Lee-Taimre-2021}, where the background intensity was assumed a constant vector with positive components.

\er

\section{The Model}

We now propose a model that describes the LOB dynamics of orders from both HFT and non-HFT investors. For the non-HFTs, we use the centered order book density model similar to that in \cite{Cont-Mueller-2021}. To model the dynamics of HFT orders, we use the multi-dimensional Hawkes process.

\ms

Let the volume of orders awaiting execution at time $t$ and price $p$ be $U(t,p)$. By convention, $U(t,p)\ges0$ for ask orders, and $U(t,p)\les0$ for bid orders. We define the ask price (the lowest ask offer) $s^a(t)$ and bid price (the highest bid offer) $s^b(t)$ as follows:
$$s^a(t):=\inf\big\{p>0,U(t,p)>0\big\},\qq s^b(t):=\sup\big\{p>0,U(t,p)< 0\big\}.$$%
We assume that all the investors are rational. Thus, they will not offer a lower price to sell than any ask price, or a higher price to buy than any bid price. Therefore,
$$s^b(t)<s^a(t),\qq\big\{U(t,p)\bigm| s^b(t)<p<s^a(t)\big\}=\varnothing.$$
With the above $s^a(t)$ and $s^b(t)$, we define the mid-price to be
$$S(t)={s^a(t)+s^b(t)\over2}.$$
We can see that
$$\ba{ll}
\ns\ds p<S(t)<s^a(t)\q\Ra\q U(t,p)\les0,\\
\ns\ds p>S(t)>s^b(t)\q\Ra\q U(t,p)\ges0.\ea$$
Let the tick size of the market be $\d>0$, and let $v(t,p)\approx U(t,p)/\d$ be the {\it volume density}. We define
$$u(t,x)=\left\{\ba{ll}
\ns\ds v(t,S(t)+x),\qq\hb{for }x\in[-L,L]\\
\ns\ds0,\qq\qq\qq\qq\hb{otherwise}.\ea\right.$$
where $L>0$, and $x$ represents a distance (either positive or negative) from the mid-price.  When $x<0$, $S(t)+x<S(t)$, and hence $u(t,x)=v(t,S(t)+x)\les0$.  Similarly, when $x>0$, $u(t,x)\ges0$. We call $u(t,x)$ the centered order book density at $(t,x)$.

\subsection{Non-HFT orders}

In this subsection, we are modeling non-HFT orders. We observe the following different LOB events with each corresponding term appeared on the right-hand side of the equation:

\ms

\it 1. Outright cancellation of orders without replacement: \rm

\ms

(i) $x>0$, $-\z_au(t,x)$, with $\z_a>0$:

\ms

This term models the decrease of $u(t,x)$ from the outright proportional cancellation of limit ask orders at the price level $S(t)+x$.

\ms

(ii) $x<0$, $-\z_bu(t,x)=\z_b|u(t,x)|$, with $\z_b>0$:

\ms

This term models the decrease of the \textit{absolute} value of $u(t,x)$ from the outright cancellation of limit bid orders at the price level $S(t)+x$.

\ss

The C-M Model \cite{Cont-Mueller-2021} contained these two terms as well with similar explanations. For notational simplicity, in what follows, we will assume $\z_a=\z_b=\z$.

\ms

\it2. Symmetric changes: \rm

\ss

(i) $x>0$: $\eta_a u_{xx}(t,x)$ with $\eta_a>0$:

\ss

This term models the symmetric changes of limit ask orders at a distance $x$ from the mid-price. For example, in the illustrative LOB (\ref{LOB}), the volume at the price level \$100.03 is lower than all the neighboring price levels, \$100.02, \$100.04, and \$100.05, which acts roughly like a local minimum and makes $u_{xx}(t,x)>0$, assuming everything is smooth. Some of the limit ask orders at the neighboring price levels will be cancelled and added to the price level \$100.03. So at the price level \$100.03, $u(t,x)$ goes up with a possible change $\eta_a u_{xx}(t,x)>0$. On the other hand, the volume at the price level \$100.02 is higher than the neighboring price levels, \$100.01, \$100.03, \$100.04 and \$100.05, which acts roughly like a local maximum and it makes $u_{xx}(t,x)<0$, assuming again everything is smooth. Some of the limit ask orders at the price level \$100.02 will be cancelled and added to the neighboring price levels. So at the price level \$100.02, $u(t,x)$ goes down with a possible change $\eta_a u_{xx}(t,x)<0$.

\ss

(ii) $x<0$: $\eta_b u_{xx}(t,x)$ with $\eta_b>0$:

\ss

This term models the symmetric changes of limit bid orders at a distance $|x|$ from the mid-price. It is similar to the ask case but applied in the opposite way since $u(t,x)<0$ by convention. For example, in the illustrative LOB (\ref{LOB}), the volume at the price level \$99.97 is lower than all the neighboring price levels, \$99.96, \$99.98, and \$99.99. Since $u(t,x)<0$, it acts roughly like a local maximum and leads to $u_{xx}(t,x)<0$, assuming everything is smooth.  Some of the limit bid orders at the neighboring price levels will be cancelled and added to the price level \$99.97. So $u(t,x)$ at the price level \$99.97 should go down with a possible change $\eta_b u_{xx}(t,x)<0$, which makes $u(t,x)$ smaller or the \textit{absolute} value $|u(t,x)|$ larger. On the other hand, the volume at the price level \$99.99 is higher than the neighboring price levels, \$99.96, \$99.97, \$99.98 and \$100. Since $u(t,x)<0$, it acts roughly like a local minimum and leads to $u_{xx}(t,x)>0$, assuming again everything is smooth.  Some of the limit bid orders at the price level \$99.99 will be cancelled and added to the neighboring price levels. So $u(t,x)$ at the price level \$99.99 should go up with a possible change $\eta_b u_{xx}(t,x)>0$, which makes $u(t,x)$ larger or the \textit{absolute} value $|u(t,x)|$ smaller.

\ms

The C-M Model \cite{Cont-Mueller-2021} also contained these two terms.  We slightly modify the notation: for example, instead of $\Delta u(t,x)$, we use $u_{xx}(t,x)$ to clarify that $x$ is one-dimensional. Again, we will assume $\eta_a=\eta_b=\eta$ for simplicity.

\ms

\it3. Cancellation of orders with asymmetric replacement: \rm

\ss

(i)  $x>0$: $-\b_a [u_x(t,x)]^-$, with $\b_a>0$:

\ss

This term models the cancellation of ask orders at a distance $x$ from the mid-price and replacement of these orders by some closer to the mid-price. When $u_x(t,x)<0$, it roughly means that there are more ask orders at lower prices than $S(t)+x$. Therefore, in order to sell the shares at the price level $S(t)+x$ faster, some investors will likely cancel their limit ask orders and resubmit them as limit ask orders at a price level closer to the mid-price, or even market ask orders. Thus, at price level $S(t)+x$, a certain portion of the volume will be decreased. This amount is assumed to be $-\beta_a [u_x(t,x)]^-$. When $u_x(t,x)>0$, it roughly means that there are more ask orders at higher prices than $S(t)+x$. Therefore, most rational investors will not cancel the orders, as.their ask orders are already better than most other orders. Hence, these orders will most likely be unchanged or the change will be $-\b_a [u_x(t,x)]^-=0$.

\ss

(ii) $x<0$: $\b_b [u_x(t,x)]^-$ with $\b_b>0$:

\ss

This term models the cancellation of bid orders at a distance $|x|$ from the mid-price and replacement of these orders closer to the mid-price. When $u_x(t,x)<0$, it roughly means that there are more bid orders at higher prices than $S(t)-|x| = S(t)+x$. Therefore, in order to buy the shares at the price level $S(t)-|x|$ faster, some investors will likely cancel their limit bid orders and resubmit them as limit bid orders at a price level closer to the mid-price, or even submit market bid orders. Thus, at price level $S(t)-|x|$, a certain portion of the volume will be decreased. This amount is assumed to be $\beta_b [u_x(t,x)]^->0$. When $u_x(t,x)>0$, it roughly means that there are more bid orders at lower prices than $S(t)-|x|=S(t)+x$.  Therefore, most rational investors will not cancel the orders as their bid orders are already better than most other orders. Hence, these orders will most likely be unchanged or the change will be $\b_b [u_x(t,x)]^-=0$.

\ms

This treatment is different from the C-M model \cite{Cont-Mueller-2021}.
We zero out the term $u_x(t,x)$ when $u_x(t,x)>0$ so that the dynamics of limit order resubmission only goes towards the middle price. This is significantly different from the usual convection in the heat transfer situation. We will assume $\b_a=\b_b=\b$ also below.

\ms

\it 4. Cancellation of orders with market order replacement: \rm

\ms

When the queues are long around the mid-price, some investors will likely cancel their previously submitted limit orders in these queues, and resubmit the orders as market orders so that their orders can get executed immediately.

\ms

For example, in the illustrative LOB (\ref{LOB}), an investor originally placed a limit ask order of 70 shares at the price level \$100.01 at 10:00 am. She wants to sell her shares relatively quickly, but she has to wait until the 3000 shares, placed before 10:00 am at the same or lower prices, to be sold first. If she wants to sell her 70 shares by 10:15 am, but does not think the 3000 shares will be sold by that time, she might cancel her order and resubmit it as a market ask order.  She would take a total loss of \$0.70, but the 70 shares can be sold immediately, executed against the existing limit bid queue at the price level \$100.00. In this case, the limit ask queue at the price level \$100.01 is decreased by 70 shares due to the cancellation, and the limit bid queue at \$100.00 is also decreased by 70 shares due to the resubmitted market bid order.

\ms

On the opposite side, another investor originally placed a limit bid order of 80 shares at the price level \$100.00 at 10:00 am. He wants to buy 80 shares relatively quickly, but he has to wait until the 2000 offers, placed before 10:00 am at same or higher prices, to be executed first. If he wants to buy 80 shares by 10:10 am, and he does not think the 2000 offers will be executed by that time, he might cancel his order and resubmit it as a market bid order. He would have to pay \$0.80 more than his previous offer, but he would get the 80 shares immediately from the existing limit ask queue at the price level \$100.01. In this case, the limit bid queue at the price level \$100.00 is decreased by 80 offers due to the cancellation, and the limit ask queue at \$100.01 is also decreased by 80 shares due to the resubmitted market ask order.

\ms

To model the impact from this LOB event, we first set a threshold $u_0>0$ such that a queue $u(t,x)$ is ``too long" if $|u(t,x)|\ges u_0$.  In other words, this LOB event will not happen when $\big|u(t,|x|)\big|<u_0$.

\ms

(i) For $x>0$, when $u(t,x)\ges u_0$, it means that the limit ask queue is too long for the investors.  Therefore, the investors that want to sell their shares of the stock quickly will likely cancel their previously submitted limit ask orders and resubmit them as market ask orders.  The cancellation will cause the limit ask volume density to decrease, and we model this impact by $-j(x)\big( u(t,x)-u_0\big)^+$ (on the right-hand side of the equation), with $j(x)$ a positive function decreasing in $x>0$, meaning that the higher a price level if above the mid-price, the less likely the investors will cancel the limit ask orders at the price level, as otherwise the loss would be too large.  Assuming all the cancelled limit ask orders become market orders, these orders will cause the \textit{absolute value} of the bid volume density to decrease, and we model this impact by $j(|x|)\big(u(t,|x|)-u_0\big)^+$. In summary, we model this scenario by
$$-j(x)\big(u(t,x)-u_0\big)^+{\bf1}_{\{x>0\}}+j(|x|)\big(u(t,|x|)-u_0\big)^+{\bf1}_{\{x<0\}}.$$

\ms

(ii) For $x<0$, it is symmetric.  When $u(t,x)\les-u_0$, it means that the limit bid queue is too long for the investors.  Therefore, the investors that want to buy the stock quickly will likely cancel their previous submitted limit bid orders and resubmit them as market bid orders.  The cancellation will cause the \textit{absolute value} of the limit bid volume density to decrease, and we model this impact by $j(x)\big(u(t,x)+u_0\big)^-$, with $j(x)$ a positive function increasing in $x<0$.  The meaning is similar to the case of $x>0$.  Also, assuming that all cancelled limit bid orders become market bid orders, then these orders will cause the limit ask volume density to decrease, and we model this impact by $-j(-|x|)\big( u(t,-|x|)+u_0\big)^-$.  In summary, we model this scenario by
$$j(x)\big(u(t,x)+u_0\big)^- \mathbf{1}_{\{x<0\}} -j(-|x|)\big( u(t,-|x|)+u_0\big)^- \mathbf{1}_{\{x>0\}}.$$
Therefore, the rate of cancellation with market order replacement at time $t$ and price level $S(t)+x$ can be modeled as
\bel{J}\ba{ll}
\ns\ds J(x,u(t,x))={\bf1}_{\{x>0\}}\big[-j(x)\big(u(t,x)-u_0\big)^+-j(-|x|)\big( u(t,-|x|)+u_0\big)^-\big]\\
\ns\ds\qq\qq\qq\qq+{\bf1}_{\{x<0\}}\big[j(|x|)\big(u(t,|x|)-u_0\big)^++j(x)\big(u(t,x)+u_0\big)^-\big]\\
\ns\ds\qq\qq\qq=-\sgn(x)\big[j(|x|)\big(u(t,|x|)-u_0\big)^++j(-|x|)\big(u(t,-|x|)+u_0\big)^-\big],\ea\ee
with $j(x)$ being a positive function decreasing in $x>0$ and increasing in $x<0$.

\ms

\it 5. Submission of orders: \rm

\ms

The submission of limit orders and market orders are both largely influenced by the price, which in turn is largely influenced by the difference between the volume of the ask and bid orders around the mid-price. We introduce
\bel{ell}\ell(t)=\int^{\iota}_{-\iota}u(t,y)dy,\ee
with $\d\les\iota\ll L$. If $\ell(t)>0$, then there are more limit ask orders than limit bid orders around the mid-price. If $\ell(t)<0$, then there are more limit bid orders than limit ask orders around the mid-price.

\ms

(i) $x>0$:

\ss

Let $\ell(t)>0$. It means that there are already too many ask orders. Therefore, rational investors are less likely to submit more limit ask orders and some investors may even cancel their previously submitted limit ask orders and wait until the ask and bid queues are balanced again. This will lead to the decreasing tendency of the limit ask orders.  Clearly, it is acceptable that the larger the $\ell(t)$, the larger the decreasing tendency. Hence, we may model this by having a term $G(x,\ell(t))$ on the right-hand side of the equation, with $G(x,\ell)$ strictly decreasing in $\ell>0$ and $G(x,0)=0$. (Thus, $G(x,\ell)<0$ if $\ell>0$.).

\ss

Let $\ell(t)<0$. It means that there are already too many bid orders, which might signal a large demand for the stock. Therefore, rational investors are more likely to submit more limit ask orders than to rush the sale with market ask orders, for a potential increase in the mid-price. This will lead to the increasing tendency of the limit ask orders.  Clearly, it is acceptable that the smaller the $\ell(t)$, the larger the increasing tendency.  Hence, we still model this by having a tern $G(x,\ell(t))$ on the right-hand side of the equation with the function $G(x,\ell)$ being strictly decreasing in $\ell$ and $G(x,0)=0$. (Thus, $G(x,\ell)>0$ if $\ell<0$.)

\ms

(ii) $x<0$, it is symmetric:

\ss

Let $\ell(t)<0$. It means that there are already too many bid orders. Thus, rational investors are less likely to submit more limit bid orders and some investors might even cancel their previous submitted limit bid orders and wait until the ask and bid queues are balanced again. This will lead to the decreasing tendency of the limit bid orders. Clearly, it is acceptable that the smaller the $\ell(t)$, the larger the decreasing tendency to the \textit{absolute value} of limit bid orders, which means the larger the increasing tendency to the bid volume density. Hence, we model this by having a term $G(x,\ell(t))$ on the right-hand side of the equation with $G(x,\ell)$ being strictly decreasing in $\ell$ and $G(x,0)=0$. (Thus, $G(x,\ell)>0$ if $\ell<0$.)

\ss

Let $\ell(t)>0$. It means that there are already too many ask orders, which might signal a large supply for the stock. Therefore, rational investors are more likely to submit limit bid orders that to rush the purchase with market bid orders, for a potential decrease in the mid-price. This will lead to the increasing tendency of the limit bid orders. Clearly, it is acceptable that the larger the $\ell(t)$, the larger the increasing tendency to the \textit{absolute value} of limit bid orders, which means the larger the decreasing tendency to the bid volume density. Hence, we still model this by having a term $G(x,\ell(t))$ on the right-hand side of the equation with $G(x,\ell)$ strictly decreasing in $\ell$ and $G(x,0)=0$. (Thus, $G(x,\ell)<0$ if $\ell>0$.)

\ms

The impact of the non-HFT order flows can be summarized by the following differential equation for the centered order book density $u$:
$$du(t,x)=\Big[\eta u_{xx}(t,x)-\b\sgn(x)[u_x(t,x)]^--\z u(t,x)+J(x,u(t,x))+G\(x,\int^{\iota}_{-\iota}u(t,y)dy\)\Big]dt,$$
where $\eta,\b,\z>0$, $J(x,u(t,x))$ is given by \rf{J}, $G(x,\ell(t))$ is described as above, i.e., $G(x,\ell)$ is strictly decreasing in $\ell$ and $G(x,0)=0$, with $\ell(\cd)$ given by \rf{ell}.

\subsection{HFT orders}

In this subsection, we are modeling the HFT orders.  We assume that the HFT orders mainly occur near the mid-price and on average they provide zero or very small net contribution in volume to the limit order book.  Thus, roughly speaking, the HFT dynamics are almost like a zero mean noise process.

\subsubsection{A microscopic HFT volume model}\label{hft-micro-model}

In order to model the volume of HFT orders on both sides of the market, we consider the following six types of market events: Submission of limit ask/bid orders, cancellation of limit ask/bid orders, and submission of market ask/bid orders. To simplify the HFT microscopic model, we reduce the dimension of our model by combining the cancellation of limit ask (bid) orders with the submission of market bid (ask) orders since their impacts on the order dynamics are the same: decrease the volume of limit ask (bid) orders. For example, by cancelling an limit ask order, it is equivalent to putting a same size market bid order since both orders are executed against the existing limit ask orders.

\ms

Viewing HFT macroscopically, it is just like a noise, and viewing it microscopically, it is mutually self-exciting. Assume the average trading speed of HFT is $n$ times per millisecond. Then, during the time interval $[0,t]$ (with $t$ being measured by second), there would be $1000nt$ trades. Thus, the number of HFTs is roughly the same as that of non-HFTs during $[0,1000t]$. Now, in general if the ratio of the fast and the slow times is $T$ (instead of $1000$), then within the (normal) time interval $[0,t]$, the average number of HFTs is roughly the same as those non-HFTs during $[0,tT]$. Hence, it is a suitable approach to investigate the HFT as follows: Consider a multi-dimensional Hawkes process $\BN_T(\cd)$, parameterized by $T\in\dbN$ (so that it is mutually exciting among components) on $[0,\i)$. Then for a sequence $T\to\i$, look at the scaling (or normalized) limit behavior of $\BN_T(\cd\,T)$, capturing the oscillation feature of it. This will be a good approximation for a model of the HFT.

\ms

Now, we make it more precise. Let $(\O^T,\cF^T,\dbP^T)$ be a complete probability space, with $T\in\dbN$ being a parameter, on which a 4-dimensional Hawkes process $\BN_T(\cd)$ is defined with $\dbF^T=\{\cF^T_t\}_{t\ges0}$ being the filtration generated by $\BN_T(\cd)$. Assume
\bel{N}\BN_T(\t)=\begin{pmatrix}
            N^{a,+}_T(\t)\\
            N^{b,+}_T(\t)\\
            N^{a,-}_T(\t)\\
            N^{b,-}_T(\t)
            \end{pmatrix},\qq\t\ges0.\ee
Here, we use $N^{a,+}_T(tT)$ ($N^{b,+}_T(tT)$) to model the accumulative number of limit ask (bid) orders submitted in the time interval $[0,tT]$, and use $N^{a,-}_T(tT)$ ($N^{b,-}_T(tT)$) to model the accumulative number of combined market ask (bid) orders and cancelled bid (ask) orders in the time interval $[0,tT]$. We will make more concrete assumptions on the above Hawkes process later on. According to the above, the accumulative volume of HFT around the mid price in the time interval $[0,tT]$ is $V_T(tT)$:
\bel{V_T}V_T(tT)=N_T^{a,+}(tT)+N_T^{b,+}(tT)-N_T^{a,-}(tT)-N_T^{b,-}(tT)
=(1,1,-1,-1)\BN_T(tT),\qq t\ges0.\ee
On the other hand, we can write the volume density $u(t,x)$ into the following generic equation:
$$u(t,x)=\hb{non-HFT volume density}+\hb{HFT volume density}.$$
Since the HFT volume density is a part of (total) volume density $u(t,x)$, we may let
$$\hb{HFT volume density}=f(t)\cd u(t,x),$$
with some function $f(t)$ valued in $(0,1)$, which serves as a ratio/fraction function in the model, so that $f(t)\cd u(t,x)$ captures the possible fluctuation of the (total) volume density due to the HFT, and preserves the same macroscopic properties of a normalized $V_T(tT)$. Such a normalization is necessary because the amplitude of $V_T(tT)$ is divergent as $T\to\i$, and only the limit of the normalized HFT volume can eventually capture the nature of the mean zero noise of HFT.  Hence, we have
\bel{f}f(t)=\lim_{T\to\i}\big[h(T)V_T(tT)\big],\ee
for some scaling factor $h(T)$. Therefore, we can model the change of HFT volume density as
$$df(t) \cdot u(t,x) + f(t)\cdot du(t,x).$$
Since the change of $u(t,x)$, observed in normal time like seconds, is significantly slower than the change of $V_T(tT)$, the impact from $du(t,x)$ to the change of HFT volume density is almost negligible. So we set $f(t)\cd du(t,x)\approx0$. Combining with the non-HFT volume density model, we have the following centered order volume density equation:
\bel{SPDE1}\ba{ll}
\ns\ds du(t,x)=\[\eta u_{xx}(t,x)-\b\sgn(x)[u_x(t,x)]^--\z u(t,x)+J(x,u(t,x))\\
\ns\ds\qq\qq\qq\qq\qq+G\(x,\int^{\iota}_{-\iota}u(t,y)dy\)\]dt+u(t,x) df(t),\ea\ee
where $V_T(\cd\,T)$ and $f(\cd)$ are given by \rf{V_T} and \rf{f}. In the next section, we will find $f(\cd)$.

\subsubsection{The Hawkes conditional intensity process}

In this subsection, we present some detailed analysis on the Hawkes process, making some hypotheses, and preparing to find its scaling limit. We drop the subscript $T\in\dbN$ as it is fixed in the subsection.

\ms

Denote the conditional intensity of the Hawkes process $\BN(\cd)$ by the following:
$$\BBl(\t)= \begin{pmatrix}
            \l^{a,+}(\t)\\
            \l^{b,+}(\t)\\
            \l^{a,-}(\t)\\
            \l^{b,-}(\t)
            \end{pmatrix},\qq\t\ges0,$$
and it is of the form
\bel{3-BBl}\BBl(\t)=\BBm(\t)+\int_0^\t\BBF(\t-s)d\BN(s)\equiv\BBm(\t)+\big[\BBF*d\BN\big](\t),\ee
where
\bel{BBm,BBF}\BBm(\cd)=\begin{pmatrix}
    \m^{a,+}\\
    \m^{b,+}\\
    \m^{a,-}\\
    \m^{b,-}
    \end{pmatrix}(\cd), \qq\BBF(\cd)=\begin{pmatrix}
    \f_{11} & \f_{12} & \f_{13} & \f_{14} \\
    \f_{21} & \f_{22} & \f_{23} & \f_{24} \\
    \f_{31} & \f_{32} & \f_{33} & \f_{34} \\
    \f_{41} & \f_{42} & \f_{43} & \f_{44}
    \end{pmatrix}(\cd).\ee
For the subscripts of each entry of $\mathbf{\Phi}(\cdot)$, $1$ stands for limit ask orders, $2$ for limit bid orders, $3$ for market ask orders, and $4$ for market bid orders.

\ms

In the conditional intensity process, $\BBm(\cd)$ models the conditional intensity that a new HFT order event is induced exogenously. For example, $\m^{a,+}(\cd)$ models the conditional intensity that a new HFT limit ask order is submitted due to some exogenous reason. The kernel matrix $\BBF(\cd)$ models the endogenous induction power from past events. For example, $\int^{\cd}_0\f_{42}(\cd-s)dN^{b,+}(s)\equiv[\f_{42}*dN^{b,+}](\cd)$ models the conditional intensity of market bid order submission induced by past limit bid order submissions.  We summarize in the following table the endogenous induction power from all the entries in $\BBF(\cd)$:

\ms

\noindent\begin{tabular}{|p{2.2cm}|p{2cm}|p{2cm}|p{2cm}|p{2cm}|p{2cm}|}
 \hline
 \multicolumn{2}{|l|}{ \multirow{2}{*}{\makecell{Conditional intensity of current order sub- \\ -mission induced by past order submission}}} & \multicolumn{4}{c|}{Current Order Submission}\\ \cline{3-6}
\multicolumn{2}{|l|}{ \multirow{2}{*}{ }} & \makecell{Limit Ask } & \makecell{Limit Bid  } & \makecell{Market Ask }  & \makecell{Market Bid }\\
 \hline
 \multirow{4}{*}{\makecell{Past Order \\ Submission} } & Limit Ask & $\f_{11} * dN^{a,+}$ & $\f_{21}*dN^{a,+}$ & $\f_{31}*dN^{a,+}$ & $\f_{41}*dN^{a,+}$\\ \cline{2-6}
 & Limit Bid & $\f_{12}*dN^{b,+}$ & $\f_{22}*dN^{b,+}$ & $\f_{32}*dN^{b,+}$ & $\f_{42}*dN^{b,+}$ \\ \cline{2-6}
 & Market Ask & $\f_{13}*dN^{a,-}$ & $\f_{23}*dN^{a,-}$ & $\f_{33}*dN^{a,-}$ & $\f_{43}*dN^{a,-}$\\ \cline{2-6}
 & Market Bid & $\f_{14}*dN^{b,-}$ & $\f_{24}*dN^{b,-}$ & $\f_{34}*dN^{b,-}$ & $\f_{44}*dN^{b,-}$\\ \cline{2-6}
 \hline
\end{tabular}\\

From the above, we see that $\f_{ij}(\cd)$ stands for the effect from past submission $j$ to current submission $i$. Now we make some analysis on the entries $\f_{ij}(\cd)$ of $\BBF(\cd)$, which will lead to some proper assumptions on these functions.

\ms

\it 1. $\f_{ii}(\cd)$, $i=1,2,3,4$. \rm

\ms

Institutional investors normally split large orders (called parent orders) into smaller orders (called children orders) and execute these smaller orders in an extended time period \cite{Almgren-Chriss-2001} \cite{Lehalle-Laruelle-2013}.  Therefore, we can assume that the conditional intensity of one type of HFT limit (market) order induced by the same type of HFT limit (market) orders in the past should be the same. In other words, the conditional intensity of the submission of one limit ask order induced by past submissions of limit ask orders can be assumed to be the same with the conditional intensity of the submission of one limit bid order induced by past submissions of limit bid orders. We let this inducing effect be a positive and bounded function $\f(\cd)$ with some $\|\f(\cd)\|_1>0$. Thus,
\bel{f_11,22}\f_{11}(\cd)=\f_{22}(\cd)=\f(\cd).\ee
As for the market (ask and bid) orders, we also assume that the conditional intensity of the submission of one market ask order induced by past submissions of market ask orders is the same with the conditional intensity of the submission of one market bid order induced by past submissions of market bid orders.  However, institutional investors tend to use limit parent orders over market parent orders, due to the lack of price control of the market parent orders \cite{Said-Ayed-Husson-Abergel-2018}. Therefore, we let the inducing effect between market orders be $\b_1\f(\cd)$ with $\b_1<1$, instead of $\f(\cd)$.

\ms

We also assume momentum effect in market orders because some individual investors want to follow the market move and they usually want to execute their orders immediately.  However, these individual investors usually do not have orders nearly as large as the ones from institutional investors, and hence this inducing power is less than that from the parent orders. We use $\beta_2 \f(\cdot)$ with $0<\beta_2<\beta_1$ to model this momentum effect.  Combining the momentum effect with the inducing effect on market orders, we should have
\bel{f_33,44}\f_{33}(\cd)=\f_{44}(\cd)=(\b_1+\b_2)\f(\cd).\ee
Note that there will be no restriction of the size on the positive number $\b_1+\b_2$, so it is possible that $\b_1+\b_2>1$, or namely, it is possible that
$$\f_{33}(\cd)=\f_{44}(\cd)>\f_{11}(\cd)=\f_{22}(\cd).$$

\it 2. $\f_{13}(\cd)$, $\f_{31}(\cd)$, $\f_{24}(\cd)$, $\f_{42}(\cd)$, $\f_{41}(\cd)$, $\f_{32}(\cd)$.

\rm

\ms

Market orders near the mid-price can potentially deplete the queues near the mid-price, which could lead to price changes, and the price changes in turn could lead to the submission of limit orders on the same side. For example, in the illustrative LOB (\ref{LOB}), if an investor places a market ask order for 2000 shares at 10:01am, the market ask order will be executed at the price level \$100.00 against the bid queue at that price level.  Since the bid queue at the price level \$100.00 only has 2000 shares, it will be depleted and the best bid price will decrease by 1 tick to \$99.99.  Meanwhile, the best ask price will also decrease by 1 tick to \$100.00.  This will likely induce the submission of limit ask orders at the new best ask price by market makers, who place limit orders at the best bid and ask prices to earn the spread. We assume that this inducing effect from market orders to limit orders on the same side is the same momentum effect between market orders, since they are both driven by price changes, so we also use $\b_2$ to model this effect, namely,
$$\f_{13}(\cd)=\f_{24}(\cd)=\b_2\f(\cd).$$
On the other hand, the high frequency limit orders signal a demand on one side, which could induce market order on the same side because speculating investors might want to act before large limit orders.  For example, if an investor observes a stable flow of incoming limit bid orders from the same institution, this could signal a parent limit bid order, which will typically take hours or even days to complete and will potentially raise the price due to the increased demand.  The investor might want to submit market bid orders so that she can buy shares of the stock at \$100.01, the current best ask price, before the potential price increase caused by the completion of this parent limit bid order.  After the entire parent limit bid order is placed, she could place a market ask order to sell these shares back to the institution at a price higher than \$100.01.  For the first step of this strategy, we assume that this inducing effect from limit orders to market orders on the same side is the same momentum effect between market orders, since they are both driven by price changes, so we still use $\beta_2$ to model this effect, namely,
$$\f_{31}(\cd)=\f_{42}(\cd)=\b_2\f(\cd).$$
As for the second step of this strategy, the investor in our example might have the wrong speculation: The flow of limit bid orders might not end up being a parent limit bid order, or the price might not increase from the sequence of limit bid orders. In this case, the investor might not submit the market ask order since it would not profit her. Therefore, we assume that the inducing effect from limit orders to market orders on the opposite side is less than that to the market orders on the same side, namely,
$$\f_{41}(\cd)<\f_{31}(\cd),\qq\f_{32}(\cd)<\f_{42}(\cd).$$
We assume $\b_3<1$ and the above can be written as
$$\f_{41}(\cd)=\b_3\f_{31}(\cd)=\b_3\b_2\f(\cd),\qq \f_{32}(\cd)=\b_3\f_{42}(\cd)= \b_3\b_2\f(\cd).$$

\ms

\it 3.  $\f_{12}(\cd)$, $\f_{21}(\cd)$, $\f_{34}(\cd)$, $\f_{43}(\cd)$ \rm

\ms

We assume that the same event on opposite sides induce each other in the same way but very close to 0.  For example, we assume that the submissions of limit ask orders barely induce the submissions of limit bid order, which is observed by the numerical results from \cite{Andersen-2014} and \cite{Bacry-Jaisson-Muzy-2016}.  So we have
\bel{f_12,34}\f_{12}(\cd)=\f_{21}(\cd)=\f_{34}(\cd)=\f_{43}(\cd)=0.\ee

\it 4. $\f_{14}(\cd)$ and $\f_{23}(\cd)$. \rm

\ms

Since we assume that the HFT orders provide almost zero net contribution in volume to the LOB on average, we have
$$\ba{ll}
\ns\ds\dbE[dV(\t)]=\dbE\big[dN^{a,+}(\t)+dN^{b,+}(\t)-dN^{a,-}(\t)-dN^{b,-}(\t)\big]\\
\ns\ds\qq\qq=\(\dbE\big[\l^{a,+}(\t)\big]+\dbE\big[\l^{b,+}(t)\big]-\dbE\big[\l^{a,-}(t)
\big]-\dbE\big[\l^{b,-}(\t)\big]\)d\t=0.\ea$$
A further careful analysis reveals that the number of limited asked orders should be roughly equal to that of market bid orders, and the number of limited bid orders should be roughly equal to that of market ask orders. Thus, it is reasonable to assume that
\bel{l=l}\ba{ll}
\ns\ns\dbE\big[\l^{a,+}(\t)\big]=\dbE\big[\l^{b,-}(\t)\big],\qq\dbE\big[\l^{b,+}(\t)\big]
=\dbE\big[\l^{a,-}(\t)\big],\\
\ns\ds\m^{a,+}(\t)=\m^{b,-}(\t),\qq\m^{b,+}(\t)=\m^{a,-}(\t),\ea\qq\t\ges0.\ee
Now,
$$\ba{ll}
\ns\ds\dbE\big[\l^{a,+}(\t)\big]=\m^{a,+}(\t)+\int^\t_0\(\f_{11}(\t-s) \dbE\big[\l^{a,+}(s)\big]+\f_{12}(\t-s)\dbE\big[\l^{b,+}(s)\big]\\
\ns\ds\qq\qq\qq\qq\qq\qq+\f_{13}(\t-s)\dbE\big[\l^{a,-}(s)\big]+\f_{14}(\t-s) \dbE\big[\l^{b,-}(s)\big]\)ds \\
\ns\ds\qq\qq\q=\m^{a,+}(\t)+\int^\t_0\(\big[\f(\t-s)+\f_{14}(\t-s)\big] \dbE\big[\l^{a,+}(s)\big]+\b_2\f(\t-s)\dbE\big[\l^{b,+}(s)\big]\)ds,\\
\ns\ds\dbE\big[\l^{b,+}(\t)\big]=\m^{b,+}(\t)+\int^\t_0\(\f_{12}(\t-s) \dbE\big[\l^{a,+}(s)\big]+\f_{22}(\t-s)\dbE\big[\l^{b,+}(s)\big]\\
\ns\ds\qq\qq\qq\qq\qq\qq+\f_{23}(\t-s)\dbE\big[\l^{a,-}(s)\big]+\f_{24}(\t-s) \dbE\big[\l^{b,-}(s)\big]\)ds\\
\ns\ds\qq\qq\q=\m^{b,+}(\t)+\int^\t_0\(\b_2\f(\t-s)\dbE
\big[\l^{a,+}(s)\big]+\big[\f(\t-s)+\f_{23}(\t-s)\big] \dbE\big[\l^{b,+}(s)\big]\)ds\\
\ns\ds\dbE\big[\l^{a,-}(\t)\big]=\m^{a,-}(\t)+\int^\t_0\(\f_{31}(\t-s) \dbE\big[\l^{a,+}(s)\big]+\f_{32}(\t-s)\dbE\big[\l^{b,+}(s)\big]\\
\ns\ds\qq\qq\qq\qq\qq\qq+\f_{33}(\t-s)\dbE\big[\l^{a,-}(s)\big]+\f_{34}(\t-s) \dbE\big[\l^{b,-}(s)\big]\)ds \\
\ns\ds\qq\qq\q=\m^{a,-}(\t)+\int^\t_0\(\b_2\f(\t-s) \dbE\big[\l^{a,+}(s)\big]+(\b_2\b_3+\b_1+\b_1+\b_2)\f(\t-s)\dbE\big[\l^{b,+}(s)\big]\)ds,\\
\ns\ds\dbE\big[\l^{b,-}(\t)\big]=\m^{b,-}(\t)+\int^\t_0\(\f_{41}(\t-s) \dbE\big[\l^{a,+}(s)\big]+\f_{42}(\t-s)\dbE\big[\l^{b,+}(s)\big]\\
\ns\ds\qq\qq\qq\qq\qq\qq+\f_{43}(\t-s)\dbE\big[\l^{a,-}(s)\big]+\f_{44}(\t-s) \dbE\big[\l^{b,-}(s)\big]\)ds \\
\ns\ds\qq\qq\q=\m^{b,-}(\t)+\int^\t_0\((\b_2\b_3+\b_1+\b_2)\f(\t-s) \dbE\big[\l^{a,+}(s)\big]+\b_2\f(\t-s)\dbE\big[\l^{b,+}(s)\big]\)ds.\ea$$
Then, the first line in \rf{l=l} is implied by the second line and
\bel{f_23,14}\f_{23}(\t)=\f_{14}(\t)=(\b_1+\b_2+\b_2\b_3-1)\f(\t)>0,\qq\t\ges0.\ee
We will keep the above assumption in the rest of the paper.

\ms

From the above analysis, we see that the conditional intensity $\BBl(\cd)$ is given by \rf{3-BBl}--\rf{BBm,BBF} with $\BBF(\cd)$ given by \rf{F}, and
\bel{phi-0}\BBF_0=\begin{pmatrix}
    1 & 0 & \b_2 & (\b_1 + \b_2+\b_2\b_3-1) \\
    0 & 1 &(\b_1+ \b_2+\b_2\b_3-1) & \b_2  \\
    \b_2  & \b_2\b_3 & (\b_1+\b_2) & 0 \\
     \b_2\b_3&  \b_2  & 0 & (\b_1+\b_2)
    \end{pmatrix}.\ee

We now explore a little more about the relation among $\b_1,\b_2,\b_3$.
We assume that the inducing power between child orders of the same parent order is much larger than the inducing power between different types of orders.  For example, an institutional investor wants to buy 50,000 shares of a stock and he uses an HFT algorithm to submit the limit bid orders sequentially.  Some individual speculators might want to submit market bid orders to buy some shares before the parent order and then submit market ask orders to sell these shares back to the institutional investor to make a profit.  A child limit bid order almost guarantees the submission of another child limit bid order since they are both a part of the same parent order, while the market bid and ask orders might not be induced by a child limit bid order, since the speculators might not foresee the parent order or believe the price will increase.  Hence, we can assume that the past limit ask order submissions are more likely to induce limit ask order submissions than they induce limit bid order submission, market ask order submission, and market bid order submission combined.
This roughly means
$$\f_{11}(\cd)>\f_{21}(\cd)+\f_{31}(\cd)+\f_{41}(\cd).$$
Similarly, we also have
$$\f_{22}(\cd)>\f_{12}(\cd)+\f_{32}(\cd)+\f_{42}(\cd),$$
$$\f_{33}(\cd)>\f_{13}(\cd)+\f_{23}(\cd)+\f_{43}(\cd),$$
$$\f_{44}(\cd)>\f_{14}(\cd)+\f_{24}(\cd)+\f_{34}(\cd).$$
These inequalities lead to the following assumption:
\bel{b}1-\b_2\b_3-\b_2>0.\ee

\ms

Further, we assume that the same side limit market order induction power is greater than the opposite side limit market induction power, which is observed by the numerical results from \cite{Andersen-2014} and \cite{Bacry-Jaisson-Muzy-2016}.  For example, limit ask order submissions are more likely induced by past market ask order submissions than past market bid order submissions. Therefore, we have $\f_{13}(\t)>\f_{14}(\t)$ and $\f_{24}(\t)>\f_{23}(\t)$. This means that we should have
\bel{b**}\b_2>\b_1+\b_2+\b_2\b_3-1>0.\ee
Combing the above observations, the assumptions on $\b_1,\b_2,\b_3$ can be summarized as follows:
\bel{b-assumptions}0<\b_2<\b_1<1,\qq0<\b_3<1,\qq\b_1+\b_2\b_3<1<\b_1+\b_2+\b_2\b_3.\ee
We will keep the above in the rest of the paper. The following gives some basic facts about the matrix $\BBF_0$.

\bp{prop-eigen} \sl {\rm(i)} The eigenvalues of $\BBF_0$ (and $\BBF_0^\top$) are given by the following:
\bel{eigenvalues}\left\{\2n\ba{ll}
\ns\ds\l_1=\b_1+2\b_2+\b_2\b_3,\\
\ns\ds\l_2=1+\b_2-\b_2\b_3,\\
\ns\ds\l_3=\b_1+\b_2\b_3,\\
\ns\ds\l_4=1-\b_2-\b_2\b_3.\ea\right.\ee
Moreover, it holds that $\l_1>\l_2>\l_3>\l_4$, which implies that $\BBF_0^\top$ is diagonalizable.

\ms

{\rm(ii)} Define
\bel{v_i}v_1=\begin{pmatrix}
       \b_2(\b_3+1)\\
       \b_2(\b_3+1)\\
        \b_1+\b_2\b_3+2\b_2-1\\
        \b_1+\b_2\b_3+2\b_2-1
        \end{pmatrix}, \q
        v_2 = \begin{pmatrix}
        -1\\
        1\\
        -1\\
        1
        \end{pmatrix}, \q
        v_3 = \begin{pmatrix}
        \b_2(\b_3-1)\\
        -\b_2(\b_3-1)\\
        -(\b_1+\b_2\b_3-1)\\
        \b_1+\b_2\b_3-1
        \end{pmatrix}, \q
        v_4=\begin{pmatrix}
        1\\
        1\\
        -1\\
        -1
        \end{pmatrix}.\ee
They are eigenvectors of $\mathbf{\Phi}_0^{\top}$ corresponding to $\lambda_1$, $\lambda_2$, $\lambda_3$, $\lambda_4$, respectively.

\ep

\it Proof. \rm Routine and lengthy calculation of eigenvalues of $\BBF_0^\top$ and the corresponding eigenvectors are carried out in the appendix. We now show the inequalities among the eigenvalues. First, $\l_1>\l_2$ is equivalent to
$$\b_1+2\b_2+\b_2\b_3>1+\b_2-\b_2\b_3\q\iff\q\b_1+\b_2+\b_2\b_3-1>-\b_2\b_3,$$
which is trivially true since $\b_1+\b_2+\b_2\b_3-1>0$. Next, $\l_2>\l_3$ is true since
$$\l_3\equiv\b_1+\b_2\b_3<1<1+\b_2-\b_2\b_3\equiv\l_2.$$
Finally, $\l_3>\l_4$ is equivalent to
$$1-\b_2-\b_2\b_3<\b_1+\b_2\b_3\q\iff\q -\b_2\b_3<\b_1+\b_2+\b_2\b_3-1,$$
which is true. \endpf

\ms

If we recall $\mathbbm{1}=(1,1,1,1)^\top\in\dbR^4$, then
\bel{1v_i=0}\mathbbm{1}^\top v_1=2\b_1+4\b_2\b_3+6\b_2-2>0;\qq\mathbbm{1}^\top v_i=0,\qq i=2,3,4.\ee
Also, we note that
\bel{v_4^T}V_T(tT)=(1,1,-1,-1)\BN_T(tT)=v_4^\top\BN_T(tT).\ee
Thus, to obtain the scaling limit of $V_T(tT)$, it suffices to find that of $\BN_T(\cd)$. In the next section, we will carry out the details for that. In the current case, similar to \rf{l_1|f|=1}, we assume the following critical unstable condition:
\bel{l_1|f|=1*}\l_1\|\f(\cd)\|_1\equiv(\b_1+2\b_2+\b_2\b_3)\|\f(\cd)\|_1=1.\ee
In what follows, we will use the notation:
\bel{v_1}v_1=(v_{1,1},v_{1,2},v_{1,3},v_{1,4})^\top,\q v_1^2=\big((v_{1,1})^2,(v_{1,2})^2,(v_{1,3})^2,(v_{1,4})^2\big)^\top.\ee
Note that all the components in $v_1$ and $v_1^2$ are strictly positive.

\br{}\rm It is seen from the above that the case we encounter is exactly the one mentioned right after Definition \ref{def.2.6}. We will see further that the most of the following procedure will work provided $\BBF(\cd)=\f(\cd)\BBF_0$ with $\|\f(\cd)\|_1<\i$ and $\BBF_0$ being a matrix such that the following conditions are satisfied:
\bel{BBF_0*}\left\{\2n\ba{ll}
\ns\ds\hb{All the entries of $\BBF_0$ are non-negative};\\
\ns\ds\hb{Matrix $\BBF_0$ has 4 distinct positive eigenvalues (Thus, $\BBF_0$ is diagonalizable)};\\
\ns\ds\hb{The critical unstable condition \rf{l_1|f|=1} (which is now \rf{l_1|f|=1*}) holds.}\ea\right.\ee
However, in the sequel, we will restrict to the case $\BBF(\cd)=\f(\cd)\BBF_0$ with $\BBF_0$ given by \rf{phi-0}, \rf{b-assumptions}, and \rf{l_1|f|=1*} is satisfied.
\er

\section{Scaling Limit of the Microscopic Volume Model}

In this section, we are going to find the scaling limit $h(T)V_T(tT)$ of the accumulative HFT volume $V_T(tT)$ (see \rf{V_T}), for a proper chosen scaling factor $h(T)$ and the resulting ratio function $f(t)$ (see \rf{f}) of the HFT density.

\subsection{An asymptotic framework and the scaling factor for the conditional intensity process}
\label{AF-framework}

Recall the complete probability space $(\O^T,\cF^T,\dbP^T)$ parameterized by $T\in\dbN$. Let $\BN_T(\cd)$ be a 4-dimensional Hawkes process with $\BN_T(0)=0$, whose filtration is $\dbF^T\equiv\{\cF^T_t\}_{t\ges0}$. The conditional intensity process $\BBl_T(\cd)$ defined by
\bel{BBL_T*}\BBl_T(t)=\lim_{h\da0}{\dbE\big[\BN_T(t+h)-\BN_T(t)\,|\,\cF^T_t\big]\over h},\qq t\ges0,\ee
which is the solution to the following integral equation:
\bel{BBl_T}\BBl_T(t)=\BBm_T(t)+\int_0^t\BBF_T(t-s)d\BN_T(s),\qq t\ges0,\ee
where $\BBm_T(\cd)$ is assumed to be an $\dbR^4$-valued (deterministic) function so that $\BBm_T(\i)=\ds\lim_{t\to\i}\BBm_T(t)$ exists, valued in the interior of the first octant (thus, its components are all strictly positive). We use $\BBm_T(\cd)$ to represent the HFT intensity induced exogenously, and use the integral term in \rf{BBl_T} to represent the HFT induced endogenously. Next, we assume the following assumption, which seems to be standard in this approach: (see \cite{Jaisson-Rosenbaum-2015}, \cite{Jaisson-Rosenbaum-2016}, \cite{El_Euch-2018}, \cite{El_Euch-Fukasawa-Rosenbaum-2018})
\bel{BBF_0**}\BBF_T(t)=a_T\f(t)\BBF_0,\qq t\ges0,\ee
where $\BBF_0$ is given by \rf{phi-0}, with conditions \rf{b-assumptions} satisfied, and $\f(\cd)$ is a strictly positive bounded integrable function so that the critical unstable condition \rf{l_1|f|=1*} holds, with $\l_1$ being the largest eigenvalue of $\BBF_0$ (see \rf{eigenvalues}). Now, we introduce the following further assumption.

\ms

{\bf(H1)} For some $\a\in({1\over2},1)$ and $\bar\m>0$, the following holds:
\bel{m_T}\BBm_T(\i)\equiv\lim_{t\to\i}\BBm_T(t)=T^{\a-1}\bar\m
\mathbbm{1}+o\big(T^{\a-1}\big),\qq T>\3n>1,\ee
where, $\mathbbm{1}\equiv(1,1,1,1)^\top\in\dbR^4$. Let $\{a_T\}_{T\in\dbN}$ be a positive sequence monotonically increasingly goes to 1 such that for some $\bar a>0$, and the same $\a$ as above
\bel{a_T}a_T=1-T^{-\a}\big[\bar a+o(1)\big],\qq T>\3n>1.\ee

\ms

Let us now explain the above hypothesis. In this paper, we consider the case that as $T\to\i$, the HFT will eventually be swept all by the endogenous orders. Thus, we assume that as $T\to\i$, $\BBm_T(\i)\to\bf0$. Also, it is natural to assume that the background intensity for all these four order types should be asymptotically equal, i.e., \rf{m_T} holds, which implies
\bel{lim m_T}\lim_{T\to\i}{\BBm_T(\i)\over\m_T}=\mathbbm{1},\qq\(\m_T=|\BBm_T(\i)|_\i\equiv\max_{1\les i\les4}\BBm_T^i(\i)=T^{\a-1}\bar\m+o(T^{\a-1})\)
\ee
Likewise, \rf{a_T} implies
\bel{lim(1-a_T)}\lim_{T\to\i}T^\a(1-a_T)=\bar a.\ee
which gives the convergence order of sequence $a_T$. The same $\a$ in \rf{m_T} and \rf{a_T} means the convergence are compatible. Among other things, we have
\bel{T(1-a)m}\lim_{T\to\i}T(1-a_T)\m_T=\bar a\bar\m,\qq\lim_{T\to\i}T^{2\a-1}{1-a_T\over\m_T}={\bar a\over\bar\m}.\ee
Since, $\a\in({1\over2},1)$, we see that
\bel{1-a_T/m_Tto0}{1-a_T\over\m_T}=T^{1-2\a}\({\bar a\over\bar\m}\)\to0.\ee
These facts will be used later.
Now, let
\bel{BM_T}\BM_T(t)=\BN_T(t)-\int_0^t\BBl_T(s)ds,\qq t\ges0.\ee
It is a martingale associated with $\BN_T(\cd)$. Similar to \rf{BMBM}, we assume
\bel{BM_TBM_T}\lan\BM_T,\BM_T\ran(t)=\int_0^t\diag\BBl_T(s)ds,\qq t\ges0.\ee
Also,
\bel{N+}\BN_T(t)=\int_0^t\BBl_T(s)ds+\BM_T(t),\qq t\ges0.\ee
Then similar to \rf{BBl(t)=} and \rf{BN(t)=}, we have
\bel{BBl_T(t)}\BBl_T(t)=\sum_{k=0}^\i\big(a_T\BBF_0\big)^k[\f^{*k}*\BBm_T](t)
+\sum_{k=1}^\i\big(a_T\BBF_0\big)^k[\f^{*k}*d\BM_T](t),\qq t\ges0,\ee
and
\bel{BN_T(t)}\BN_T(t)=\sum_{k=0}^\i\big(a_T\BBF_0\big)^k\int_0^t[\f^{*k}*\BBm_T](r)dr
+\sum_{k=0}^\i\big(a_T\BBF_0\big)^k\int_0^t[\f^{*k}*d\BM_T](r)dr,\qq t\ges0.\ee
Further, similar to \rf{EBBl} and \rf{EBN},
\bel{EBBl*}\|\dbE[\BBl_T(\cd)]\|_\i\les{C\|\BBm_T(\cd)\|_\i\over1-a_T}\les{C\m_T\over1-a_T},\ee
and
\bel{EBN*}0\les\dbE[\BN_T(t)]\les{C\|\BBm_T(\cd)\|_\i t\over1-a_T}\les{C\m_T\over1-a_T}t,\qq t\ges0.\ee

Now, we want to find a proper scaling factor for $\BBl_T(\cd)$. To this end, we take expectation of \rf{BBl_T(t)} to obtain
\bel{El_T}\dbE[\BBl_T(tT)]=\sum_{k=0}^\i\big(a_T\BBF_0\big)^k[\f^{*k}*\BBm_T](tT),\qq t>0.\ee
Let $t\to\i$, by the dominated convergence theorem and monotone convergence theorem, we have
\bel{4.5}\ba{ll}
\ns\ds\lim_{t\to\i}\dbE\big[\BBl_T(tT)\big]=\lim_{t\to\i}\int_0^{tT} \sum_{k=0}^\i\big(a_T\BBF_0\big)^k\f^{*k}(s)\BBm_T(tT-s)ds\\
\ns\ds=\[\sum_{k=0}^\i\big(a_T\BBF_0\big)^k\int_0^\i\f^{*k}(s)ds\]\BBm_T(\i)\\
\ns\ds=\[\sum_{k=0}^\i\big(a_T\BBF_0\big)^k\(\int_0^\i\f(s)ds\)^k\]\BBm_T(\i)
=\[\sum_{k=0}^\i\big(a_T\|\f(\cd)\|_1\BBF_0\big)^k\]\BBm_T(\i)\\
\ns\ds=\BP\begin{pmatrix}
        {1\over1-a_T} & 0 & 0 & 0 \\
        0 & {1\over1-a_T\l_2\|\f(\cd)\|_1} & 0 & 0 \\
        0 & 0 & {1\over1-a_T\l_3\|\f(\cd)\|_1} & 0 \\
        0 & 0 & 0 & {1\over1-a_T\l_4\|\f(\cd)\|_1}\end{pmatrix}\BP^{-1}\BBm_T(\i).\ea\ee
The last equality is due to the facts that $0<a_T\l_i\|\f(\cd)\|_1<1$ and the critical unstable condition $\l_1\|\f(\cd)\|_1=1$ (see \rf{l_1|f|=1*}). Hence, it follows that
\bel{4.5*}\lim_{T\to\i}{1-a_T\over\m_T}\lim_{t\to\i}\dbE\big[\BBl_T(tT)\big]=\BP\begin{pmatrix}
        1 & 0 & 0 & 0 \\
        0 & 0 & 0 & 0 \\
        0 & 0 & 0 & 0 \\
        0 & 0 & 0 & 0\end{pmatrix}\BP^{-1}\mathbbm{1}.\ee
This suggests that ${1-a_T\over\m_T}\equiv T^{1-2\a}{\bar a\over\bar\m}+o(T^{1-2\a})\to0$ (since $\a\in({1\over2},1)$) should be a suitable scaling factor for $\BBl_T(tT)$. Note that although the factor ${1-a_T\over\m_T}\to0$, the above limit shows that multiplying such a factor to $\BBl_T(tT)$, it will have a non-trivial limit (neither zero nor infinite).

\ms

Based on the cluster representation of Hawkes process, for given $T\in\dbN$ (with $0<a_T<1$), the largest eigenvalue of each $\int^\i_0\BBF_T(s)ds$ can be used to model the percentage of endogenous orders in the HFT market. Thus in our model, the HFT market gets more and more endogenous over the time.

\ms

\subsection{Scaling limit of conditional intensity process}

To model the effect of a submitted order to the right after order submissions, we can choose the excitation function $\f(\cd)$ suitably. If $\f(\cd)$ is close to the Dirac delta function, it would mean that an order is unlikely to excite other orders right after it arrives at the HFT market. Thus, such a function is definitely not a suitable choice. An exponential function like $ae^{-bt}$ with $a,b>0$, could be a choice for the Hawkes excitation function. It yields an exponentially decaying conditional intensity, which means the effect of a submitted order to the later submissions only has a very short time period. This does not describe the real situation. We mentioned in the introduction, many HFT orders are part of a larger parent order that typically takes hours or even days to fully execute, which can be observed in an HFT market by child orders exciting each other during this relatively long execution window. Therefore, we model this long-term influence by assuming $\f(\cd)$ a power-law tail. More precisely, we assume the following:

\ms

{\bf(H2)} The function $\f(\cd)$ is positive, bounded, integrable, with $\l_1\|\f(\cd)\|_1=1$, and
\bel{1-int}\lim_{t\to\i}t^\a\int_t^\i\f(s)ds=\k,\ee
with the same $\a\in({1\over2},1)$ appeared in (H1), and some positive constant $\k$.

\ms

The above gives the speed of convergence of $\int_0^t\f(s)ds\to\|\f(\cd)\|_1$ as $t\to\i$. The following gives the scaling limit of $\BBl_T(tT)$.

\bp{lim-l_T} \sl Let {\rm(H1)--(H2)} hold. Then
\bel{lim-L_T*}\left\{\2n\ba{ll}
\ns\ds\lim_{T\to\i}v_i^\top\BBl_T(tT)=0,\qq t>0,~i=2,3,4,\\
\ns\ds\lim_{T\to\i}{1-a_T\over\m_T}v_1^\top\BBl_T(tT)=Y(t),\qq t>0,\ea\right.\ee
with $Y(\cd)$ being the solution to the following integral equation:
\bel{Y}Y(t)=v_1^\top\mathbbm{1}\int_0^tf^{\a,\bar\n}(t-s)ds+\bar\k\int_0^tf^{\a,\bar\n}(t-s)
\sqrt{Y(s)}\,dB_1(s),\qq t\ges0,\ee
where
\bel{f^an}\left\{\2n\ba{ll}
\ds f^{\a,\n}(t)\equiv\n t^{\a-1}E_{\a,\a}(-\n t^\a),\qq t\ges0,\\
\ns\ds E_{\a,\b}(z)=\sum_{n=0}^\i{z^n\over\G(\a n+\b)},\qq z\in\dbC,~\a,\b>0,\ea\right.\ee
and
\bel{4.24}\bar\n={\bar a\over\l_1\k\G(1-\a)},\qq\bar\k={1\over\sqrt{\bar a\bar\m}}
\sqrt{\mathbbm{1}^\top v_1^2\over\mathbbm{1}^\top v_1}.\ee

\ep

\ms
In the above, $E_{\a,\b}(z)$ is called the {\it Mittag-Leffler function}, and $f^{\a,\n}(t)$ is called {\it Mittag-Leffler density function}. See \cite{Haubold-Mathai-Saxena-2011}, \cite{Gorenflo-Kilbas-Mainardi-Rogosin-2014} for more details.

\ms

\it Proof. \rm We let $\{\e_1,\e_2,\e_3,\e_4\}$ be an orthonormal basis of $\dbR^4$ with
\bel{e_i}\e_1={1\over2}\mathbbm{1},\q\hb{span}\,\{\e_2,\e_3,\e_4\}=\hb{span}\,
\{v_2,v_3,v_4\}\perp\e_1.\ee
By \rf{1v_i=0}, the above is possible. Thus, one may assume
\bel{e_i*}\e_i=\g_{i2}v_2+\g_{i3}v_3+\g_{i4}v_4,\qq i=2,3,4.\ee
for some constants $\g_{ij}$. Also, let
\bel{v'}v'=\e_1-{1\over\e_1^\top v_1}v_1.\ee
Then, $\e_1^\top v'=0$, or $v'\in\hb{span}\,\{\e_2,\e_3,\e_4\}\equiv\hb{span}\,\{v_2,v_3,v_4\}$. Hence, we may assume
\bel{v'*}v'=\g'_2v_2+\g'_3v_3+\g'_4v_4,\qq\e_1={1\over\e_1^\top v_1}v_1+v'={1\over\e_1^\top v_1}v_1+\g_2'v_2+\g_3'v_3+\g_4'v_4.\ee
Now, decomposing $\BBl_T(t)$ according to the basis $\{\e_1,\e_2,\e_3,\e_4\}$, it follows that
\bel{CT-expression}\ba{ll}
\ns\ds\BBl_T(tT)=\sum_{i=1}^4\(\e_i^\top\BBl_T(tT)\)\e_i\\
\ns\ds\qq\q=\2n\[\({1\over\e_1^\top v_1}v_1\1n+\1n\g_2'v_2\1n+\1n\g_3\1n'\1nv_3\1n+\1n\g_4'v_4\)^\top\2n\BBl_T(tT)\]\e_1
+\sum_{i=2}^4\[\(\g_{i2}v_2^\top\1n+\1n\g_{i3}v_3^\top\1n+\1n\g_{i4}v_4^\top\)
\BBl_T(tT)\]\e_i.\ea\ee
For each $i=1,2,3,4$ and for any $k\ges1$, one has
\bel{vi-CT}\ba{ll}
\ns\ds v_i^\top\BBl_T(tT)=v_i^\top\BBm_T(tT)+v_i^\top\sum_{k=1}^\i(a_T\BBF_0)^k[ \f^{*k}*\BBm_T](tT)+v_i^\top\sum_{k=1}^\i(a_T\BBF_0)^k[\f^{*k}*d\BM_T](tT)\\
\ns\ds\qq\qq=v_i^\top\BBm_T(tT)+\sum_{k=1}^\i(a_T\l_i)^k[ \f^{*k}*(v_i^\top\BBm_T)](tT)+\sum_{k=1}^\i(a_T\l_i)^k\big[\f^{*k}*d\big(v_i^\top
\BM_T\big)\big](tT)\\
\ns\ds\qq\qq\equiv v_i^\top\BBm_T(tT)+\big[\psi_{T,i}*(v_i^\top\BBm_T)\big](tT)+\big[\psi_{T,i}*d\big(v_i^\top
\BM_T\big)\big](tT)\\
\ns\ds\qq\qq=v_i^\top\BBm_T(tT)+\int_0^{tT}\psi_{T,i}(tT-s)(v_i^\top\BBm_T(s))ds+\int_0^{tT}
\psi_{T,i}(tT-s)d(v_i^\top\BM_T(s)),\ea\ee
where
\bel{psi}\psi_{T,i}(t)=\sum_{k=1}^\i(a_T\l_i)^k\f^{*k}(t),\qq t\ges0,\ee
which is deterministic. Let the Fourier transform of $\psi_{T,i}(T\,\cd)$ be $\wt\psi_{T,i}(T\,\cd)$. Then for $\t\in\dbR$, one has
\bel{hpsi_i}\ba{ll}
\ns\ds\wt\psi_{T,i}(T\,\cd)(\t)=\int_{-\i}^\i e^{-i\t s}\psi_{T,i}(Ts)ds=\sum_{k=1}^\i(a_T\l_i)^k\int_0^\i e^{-i\t s}\f^{*k}(Ts)ds\\
\ns\ds\qq\qq\q=\sum_{k=1}^\i(a_T\l_i)^k[\wt\f(T\,\cd)(\t)]^k=\sum_{k=1}^\i(a_T\l_i)^k
\(\int_0^\i e^{-i\t s}\f(Ts)ds\)^k\\
\ns\ds\qq\qq\q=\sum_{k=1}^\i(a_T\l_i)^k{1\over T}\(\int_0^\i\f(s')e^{-i\t{s'\over T}}ds'\)^k={1\over T}\sum_{k=1}^\i\[a_T\l_i\wt\f\({\t\over T}\)\]^k.\ea\ee
Now, since
\bel{|hf|}\Big|\wt\f\({\t\over T}\)\Big|\les\int_0^\i\f(s)ds=\|\f(\cd)\|_1,\ee
we have the absolute convergence of the series in \rf{hpsi_i}, and
$$\wt\psi_{T,i}(T\,\cd)(\t)={1\over T}{\ds a_T\l_i\wt\f\({\t\over T}\)\over\ds1-a_T\l_i\wt\f\({\t\over T}\)}.$$
This leads to
$$\sup_{\t\in\dbR}|\wt\psi_{T,i}(T\,\cd)(\t)|\les{1\over T}{a_T\l_i\|\f(\cd)\|_1\over1-a_T\l_i\|\f(\cd)\|_1}.$$
Next, we claim that
\bel{limv_il=0}\lim_{T\to\i}\dbE[v_i^\top\BBl_T(tT)]=0,\qq t>0,
\q i=2,3,4.\ee
In fact, for $i=2,3,4$, since $\l_i\|\f(\cd)\|_1<1$, one has
$$\lim_{T\to\i}\wt\psi_{T,i}(T\,\cd)(\t)=0,$$
uniformly in $\t\in\dbR$. By Plancherel theorem ($\2n$\cite{Serov-2017}, p.146), it follows that
$$\|\psi_{T,i}(T\,\cd)\|_2={1\over\sqrt{2\pi}}\|\wt\psi_{T,i}(\cd)\|_2\to0,\qq T\to\i.$$
Clearly, for $t>0$, when $T$ is large, noting $v_i^\top\mathbbm{1}=0$ for $i=2,3,4$, we have
$$v_i^\top\BBm_T(tT)=\[v_i^\top\BBm_T(\i)+o(1)\]=v_i^\top\[\bar\m \big(\mathbbm{1}+o(T^{\a-1})\big)+o(1)\]=o(1).$$
By the boundedness of $\BBm_T(\cd)$, we have
$$\ba{ll}
\ns\ds[\psi_{T,i}*(v_i^\top\BBm_T)](tT)=\int_0^{tT}\psi_{T,i}(tT-s)v_i^\top\BBm_T(s)ds
\qq(s=s'T)\\
\ns\ds={1\over T}\int_0^t\psi_{T,i}((t-s')T)v_i^\top\BBm_T(s'T)ds'\to0,\qq\hb{as }T\to\i.\ea$$
Finally, we have (by \rf{psi*dM})
\bel{E|psi*dBM|}\dbE\big|[\psi_{T,i}*d(v_i^\top\BM_T)](tT)\big|\les\({\|v_i^\top\BBm_T(\cd)\|_\i\over1-\l_i\|\f(\cd)
\|_1}\)^{1\over2}\|\psi_{T,i}(\cd)\|_2\to0,\qq\hb{as $T\to\i$}.\ee
Thus, claim \rf{limv_il=0} holds. Consequently, \rf{CT-expression} becomes
\bel{CT-expression*}\BBl_T(tT)=\({1\over\e_1^\top v_1}v_1^\top\BBl_T(tT)\)\e_1
+o(1).\ee
Next, we define
\bel{eqn-B-T}B_{T,1}(t)={1\over\sqrt T}\int_0^{tT}{d(v_1^\top\BM_T)(s)\over\sqrt{ (v_1^2)^\top\BBl_T(s)}},\ee
recalling \rf{v_1}. Note that the components of $v_1^2$ and $\BBl_T(t)$ are all positive. Thus, the above definition make sense (The term under the squre root is positive). Then
\bel{dB_Ti}dB_{T,1}(t)={d(v_1^\top\BM_T)(tT)\over
\sqrt{T(v_1^2)^\top\BBl_T(tT)}}.\ee
Note that
$$(v_1^2)^\top\BBl_T(t)=v_1^\top\diag[\BBl_T(t)]v_1.$$
Thus,
$$\dbE\big[B_{T,1}(t)^2\big]=\dbE\Big|\int_0^{tT}{d(v_1^\top\BM_T)(sT)\over
\sqrt{T(v_1^2)^\top\BBl_T(sT)}}\Big|^2=\dbE\int_0^{tT}{v_1^\top\diag\BBl_T(sT)v_1\over T(v_1^2)^\top\BBl_T(sT)}ds={1\over T}\int_0^{tT}ds=t.$$
Therefore, the limit of $B_{T,1}(\cd)$ as $T\to\i$ is a Brownian motion (see
\cite{El_Euch-Fukasawa-Rosenbaum-2018}, p.254).

\ms

Now, we observe the case $i=1$. First of all, we need to deal with the expression under the radical sign in \rf{dB_Ti}. To this end, we decompose $v_1^2=\big((v_{1,1})^2,(v_{1,2})^2,(v_{1,3})^2,(v_{1,4})^2\big)^\top$ according to the basis $\{\e_1,\e_2,\e_3,\e_4\}$ (see \rf{e_i}) to get
$$\ba{ll}
\ns\ds v_1^2=[\e_1^\top(v_1^2)]\e_1+[\e_2^\top(v_1^2)]\e_2+[\e_3^\top(v_1^2)]\e_3+[\e_4^\top
(v_1^2)]\e_4\\
\ns\ds \ea$$
Thus, by \rf{CT-expression}, we have
$$(v_1^2)^\top\BBl_T(tT)={\e_1^\top v_1^2\over\e_1^\top v_1}v_1^\top\BBl_T(tT)+o(1).$$
By \rf{vi-CT}, we have
\bel{4.16}\ba{ll}
\ns\ds Y_T(t)\equiv{1-a_T\over\m_T}v_1^\top\BBl_T(tT)={1-a_T\over\m_T}(v_1^\top\BBm_T(tT))+\int_0^{tT}{1-a_T\over\m_T} \psi_{T,1}(tT-s)(v_1^\top\BBm_T(s))ds\\
\ns\ds\qq\qq\qq\qq\qq+\int_0^{tT}{1-a_T\over\m_T}\psi_{T,1}(tT-s)d(v_1^\top \BM_T)(s))\qq(\hb{let $s=s'T$})\\
\ns\ds\qq\q={1-a_T\over\m_T}(v_1^\top\BBm_T(tT))+\int_0^t{T(1-a_T)\over     \m_T}\psi_{T,1}((t-s')T)(v_1^\top\BBm_T(s'T))ds'\\
\ns\ds\qq\qq\qq\qq\qq+{1-a_T\over\m_T}\int_0^t\psi_{T,1}((t-s')T)d(v_1^\top \BM_T(s'T)).\ea\ee
Clearly,

$$\ba{ll}
\ns\ds{1-a_T\over\m_T}\int_0^t\psi_{T,1}((t-s')T)(v_1^\top d\BM_T(s'T))\\
\ns\ds=\sqrt{{T(1-a_T)\over \m_T}}\int_0^t\psi_{T,1}((t-s')T)\sqrt{{1-a_T\over\m_T}(v_1^2)^\top \BBl_T(s'T)}\, dB_{T,1}(s').\ea$$
Hence, \rf{4.16} becomes
\bel{4.18}\ba{ll}
\ns\ds Y_T(t)\equiv{1-a_T\over\m_T}v_1^\top\BBl_T(tT)={1-a_T\over\m_T}(v_1^\top\BBm_T(tT))
+\int_0^t{T(1-a_T)\over\m_T}\psi_{T,1}((t-s')T)(v_1^\top\BBm_T(s'T))ds'\\
\ns\ds\qq\qq\qq\qq\qq+\sqrt{{T(1-a_T)\over\m_T}}\int_0^t\psi_{T,1}((t-s')T) \sqrt{{1-a_T\over\m_T}(v_1^2)^\top\BBl_T(s'T)}\,dB_{T,1}(s').\ea\ee
Next, we examine the asymptotic behavior of $\psi_{T,1}(\cd)$, (rather than $\psi_{T,i}(\cd)$, $i=2,3,4$). Let the Laplace transform of $\psi_{T,1}(T\,\cd)$ be $\h\psi_{T,1}(T\,\cd)$. Then, similar to \rf{hpsi_i}, we have for $z\in\dbC$ with $\Re z>0$,
\bel{hpsi_1}\h\psi_{T,1}(T\,\cd)(z)=\int_0^\i e^{-zs}\psi_{T,1}(s)ds={1\over T}\sum_{k=1}^\i\[a_T\l_1\h\f\({z\over T}\)\]^k={\ds a_T\l_1\h\f\({z\over T}\)\over\ds T\[1-a_T\l_1\h\f\({z\over T}\)\]}.\ee
The series is convergent since similar to \rf{|hf|}, the following holds:
$$\Big|\h\f\({z\over T}\)\Big|\les\int_0^\i e^{-{\Re z\over T} s}\f(s)ds\les\int_0^\i\f(s)ds=\|\f(\cd)\|_1,$$
and $0<a_T<1$. Now by \rf{1-int},
$$\ba{ll}
\ns\ds\h\f\({z\over T}\)=\int_0^\i\f(s)e^{-s{z\over T}}ds=-\int_0^\i e^{-s{z\over T}}d\(\int_s^\i\f(r)dr\)ds\\
\ns\ds\qq\q=-e^{-s{z\over T}}\(\int_s^\i\f(r)dr\)\Big|_0^\i-{z\over T}\int_0^\i e^{-s{z\over T}}\(\int_s^\i\f(r)dr\)ds.\\
\ns\ds\qq\q=\|\f(\cd)\|_1-\int_0^\i e^{-s'}\(\int_{s'{T\over z}}^\i\f(r)dr\)ds'\qq\qq(\hb{set $\ds s{z\over T}=s'$})\\
\ns\ds\qq\q=\|\f(\cd)\|_1-\int_0^\i e^{-s'}\big[\k+o(1)\big]\(s'{T\over z}\)^{-\a}ds'\\
\ns\ds\qq\q=\|\f(\cd)\|_1-\({z\over T}\)^\a\big[\k+o(1)\big]\int_0^\i
e^{-s'}(s')^{-\a}ds'\\
\ns\ds\qq\q\equiv\|\f(\cd)\|_1-\({z\over T}\)^\a\big[\k+o(1)\big]\G(1-\a).\ea$$
Thus, noting \rf{a_T}, one has
\bel{ML1}\ba{ll}
\ns\ds T(1-a_T)\h\psi_{T,1}(T\,\cd)(z)={\ds(1-a_T)a_T\l_1\[\|\f(\cd)\|_1-\({z\over T}\)^\a[\k+o(1)]\G(1-\a)\]\over\ds 1-a_T\l_1\[\|\f(\cd)\|_1-\({z\over T}\)^\a\big[\k+o(1)\big]\G(1-\a)\]}\\
\ns\ds\qq\qq\qq\qq\qq={(1-a_T)a_T\(T^\a-\l_1z^\a[\k+o(1)]\G(1-\a)]\)\over
T^\a(1-a_T)+a_T\l_1z^\a\big[\k+o(1)\big]\G(1-\a)}\\
\ns\ds\qq\qq\qq\qq\qq={T^\a(1-a_T)a_T-(1-a_T)a_T\l_1z^\a[\k+o(1)]\G(1-\a)]\over
T^\a(1-a_T)+a_T\l_1z^\a\big[\k+o(1)\big]\G(1-\a)}\\
\ns\ds\qq\qq\qq\qq\qq={a_T\n_T+o(z^\a)\over\n_T+z^\a},\qq\qq\qq T>\3n>1,\ea\ee
where
\bel{nto}\n_T={(1-a_T)T^\a\over a_T\l_1[\k+o(1)]\G(1-\a)}\to{\bar a\over\l_1\k\G(1-\a)}\equiv\bar\n,\qq\hb{as }T\to\i.\ee
By \cite{Gorenflo-Kilbas-Mainardi-Rogosin-2014} (p.85), (see also \cite{Jaisson-Rosenbaum-2016} (p.2866) and \cite{El_Euch-2018} (p.50)), we know that ${\n_T\over\n_T+z^\a}$ is equal to the Laplace transform of the function $f^{\a,\n_T}(\cd)$. Therefore, by the continuity of the inverse Laplace transform (see \cite{Feller-1971}, p.431; see also \cite{Jaisson-Rosenbaum-2016} and \cite{El_Euch-2018}), we have
\bel{4.42}T(1-a_T)\psi_{T,1}(Tt)=a_T\n_Tt^{\a-1}E_{\a,\a}(-\n_Tt^\a)+o(1)\equiv a_Tf^{\a,\n_T}(t)+o(1).\ee
Plugging this back in the equation \eqref{4.18}, we get (for $\a>{1\over2}$)
\bel{(1-a_T)}\ba{ll}
\ns\ds Y_T(t)={1-a_T\over\m_T}v_1^\top\BBm_T(tT)+a_T \int_0^t(t-s)^{\a-1}[f^{\a,\n_T}(t-s)+o(1)]v_1^\top{\BBm_T(Ts)\over\m_T}ds\\
\ns\ds\qq\qq\qq+{a_T\over\sqrt{T(1-a_T)\m_T}}\int_0^t(t-s)^{\a-1}[f^{\a,\n_T}(t-s)+o(1)]
\sqrt{{1-a_T\over\m_T}(v_1^2)^\top\BBl_T(sT)}\,dB_{T,1}(s).\ea\ee
Now, let $T\to\i$, we see that the limit $Y(\cd)$ of $Y_T(\cd)$ should satisfy \rf{Y}, where $f^{\a,\bar\n}(t)$ and $\bar\n,\bar\k$ are given by \rf{f^an} and \rf{4.24}. \endpf

\ms

The above equation \rf{Y} involves function $f^{\a,\bar\n}(\cd)$ which is complicated. We desire to get a better equation form. This can be done, by a result from \cite{El_Euch-Fukasawa-Rosenbaum-2018} pp.274--275.

\bp{} \sl Process $Y(\cd)$ is a solution of \rf{Y} if and only if it is a solution to the following:
\bel{Y*}Y(t)={\bar\n\over\G(\a)}\int_0^t(t-s)^{\a-1}\big(v_1^\top\mathbbm{1}-Y(s)\big)ds
+{\bar\k\bar\n\over\G(\a)}\int_0^t(t-s)^{\a-1}\sqrt{Y(s)}dB_1(s),\q t\ges0.\ee
Moreover, both \rf{Y} and \rf{Y*} admit a unique solution.

\ep

\rm The fractional Brownian motion $B^H(t)$ can be expressed as
$$B^H(t)={1\over\G(H+{1\over2})}\(\int_0^t(t-s)^{H-{1\over2}}dW(s)+\int_{-\i}^0
(t-s)^{H-{1\over2}}-(-s)^{H-{1\over2}}dW(s)\),$$
where $W(t)$ is a Brownian motion, and $H$ is the Hurst parameter associated with $B^H(t)$ \cite{Mandelbrot-Van_Ness-1968}. Therefore, we can interpret \rf{Y*} as a Volterra integral equation with Hurst parameter $\a-{1\over2}\in(0,{1\over2})$.

\subsection{The auxiliary processes and their scaling limits}

To proceed further, we now define the following auxiliary processes:
\bel{L_T(t)}\ba{ll}
\ns\ds\BBL_T(t)={1-a_T\over T\m_T}\int_0^{tT}\BBl_T(r)dr\\
\ns\ds\qq\q={1-a_T\over T\m_T}\[\sum_{k=0}^\i\big(a_T\BBF_0\big)^k\int_0^{tT}[\f^{*k}*\BBm_T]
(r)dr+\sum_{k=1}^\i\int_0^{tT}\big(a_T\BBF_0\big)^k[\f^{*k}*d\BM_T](r)\],\ea\ee
\bel{BX_T(t)}\ba{ll}
\ns\ds\BX_T(t)={1-a_T\over T\m_T}\BN_T(tT)={1-a_T\over T\m_T}\[\int_0^{tT}\BBl_T(r)dr+\BM_T(tT)\]=\BBL_T(t)+{1-a_T\over T\m_T}\BM_T(tT)\\
\ns\ds\qq\q={1-a_T\over T\m_T}\[\sum_{k=0}^\i\big(a_T\BBF_0\big)^k\int_0^{tT}[\f^{*k}*\BBm_T](r)dr
+\sum_{k=0}^\i\int_0^{tT}\big(a_T\BBF_0\big)^k[\f^{*k}*d\BM_T](r)\],\ea\ee
and
\bel{Z_T(t)}\BZ_T(t)=\sqrt{T\m_T\over1-a_T}\(\BX_T(t)-\BBL_T(t)\)
=\sqrt{1-a_T\over T\m_T}\BM_T(tT).\ee
Clearly, each component of $\BX_T(\cd)$ is non-decreasing and non-negative; each component of $\BBL_T(\cd)$ is strictly increasing and strictly positive; and $\BZ_T(\cd)$ is a martingale. Roughly speaking, since ${1-a_T\over\m_T}$ is suggested to be a scaling factor of $\BBl_T(t)$, a scaling factor of $\int_0^{tT}\BBl_T(s)ds$ could be ${1-a_T\over T\m_T}$.

\ms

To study the convergence of the above-defined auxiliary processes, let us recall a result of Kurtz (\2n\cite{Kurtz-1991}) first. Let $M[0,\i)$ be the space of (equivalence class of) $\dbR$-valued, Borel measurable functions topologized by convergence in Lebesgue measure.\footnote{As indicated in \cite{Meyer-Zheng-1984} (see \cite{Jacod-Shiryaev-1987}, Chapter VI, also) that this topology is much weaker than the Skorohod topology. It is known that $f_n(\cd)$  converges to $f(\cd)$ in Lebesgue measure on $[0,T]$ if and if $\int_0^T{|f_n(t)-f(t)|dt\over1+|f_n(t)-f(t)|}$ goes to zero.} A process $X:[0,\i)\times\O\to\dbR$ can be regarded as $M[0,\i)$-valued random variable. Suppose a process $X(\cd)$ is adapted to some fixed filtration $\dbF=\{\cF_t\}_{t\ges0}$. We define
\bel{V_t}V_t(X(\cd))=\sup\dbE\[\sum_i\big|\dbE[X(t_{i+1})-X(t_i)|\cF_{t_i}]\big|\],\qq t>0,\ee
where the supremum is taken over all possible partitions of $[0,t]$. This is called {\it conditional variation}. When $X(\cd)$ is monotone, then
\bel{V_t*}V_t(X(\cd))=\dbE|X(t)|,\qq\forall t>0.\ee

The following result is quoted from \cite{Kurtz-1991}, pp.1033--1034, with small notational changes.

\bt{Kurtz} \sl Let $\{X_n(\cd)\}$ be a sequence of c\`adl\`ag, real-valued processes such that the following so-called Meyer-Zheng's condition holds:
\bel{MZ}C(t)\equiv\sup_{n\ges1}\(V_t(X_n(\cd))+\dbE\big[|X_n(t)|\big]\)<\i\qq\forall t>0.\ee
Then $\{X_n(\cd)\}$ is relatively compact in $M[0,\i)$, i.e., there exists a subsequence $X_{n_k}(\cd)$ convergent in $M[0,\i)$, almost surely, and any limit point $X(\cd)$ has a c\`adl\`ag version satisfying
\bel{MZ*}V_t(X(\cd))+\dbE[X(t)]\les C(t),\qq\forall t>0.\ee
In particular, if for each $n\ges1$, $X_n(\cd)$ is non-negative and nondecreasing, then the above conclusion holds only if
\bel{EX<i}0\les\sup_{n\ges1}\dbE[X_n(t)]<\i,\qq t>0.\ee

\et

Now, we have the following result for our case.

\bp{prop3.1} \sl The sequence $(\BX_T(\cd),\BBL_T(\cd),\BZ_T(\cd))$ satisfy Meyer-Zheng's condition \rf{MZ},
\bel{X-L}\lim_{T\to\i}\|\BBL_T(\cd)-\BX_T(\cd)\|_\i=0,\qq\as\ee
Moreover, if $(\BX(\cd),\BZ(\cd))$ is a limit point of $(\BX_T(\cd),\BZ_T(\cd))$, then it is continuous and $\BZ(\cd)$ is a martingale with
\bel{L=X}\BBL(t)=\BX(t)={1\over\e_1^\top v_1}\(\int^t_0 Y(s)ds\)\e_1,\qq t\ges0,\ee
with $Y(\cd)$ being the solution of \rf{Y}. Further,
\bel{Z*}\BZ(t)={1\over\sqrt{\e_1^\top v_1}}\int_0^t\sqrt{Y(s)}dB(s),\qq t\ges0,\ee
for a 4-dimensional standard Brownian motion $B(\cd)$ and consequently,
\bel{ZZ}\lan\BZ,\BZ\ran(t)=\diag\BX(t)=\diag\BBL(t)={1\over\e_1^\top v_1}\(\int^t_0 Y(s)ds\)I.\ee

\ep

\it Proof. \rm Since $\BN_T(\cd)$ is a Hawkes process, its components are non-decreasing. So is $\BBL_T(\cd)$. By \rf{EBN*}, we have
$$0\les\dbE[\BBL_T(t)]=\dbE[\BX_T(t)]={1-a_T\over
T\m_T}\dbE[\BN_T(tT)]\les{C\|\BBm_T(\cd)\|_\i t\over1-a_T}\les{C\m_T\over1-a_T}t,\qq t\ges0$$
This implies that $\BX_T(\cd)$ and $\BBL_T(\cd)$ satisfy Meyer-Zheng's condition, as their each component is monotone non-decreasing\footnote{It was claimed in \cite{Jaisson-Rosenbaum-2016}, p.2872, \cite{El_Euch-Fukasawa-Rosenbaum-2018}, p.270.  \cite{El_Euch-2018}, p.91, that these processes are $C$-tight, in the sense of \cite{Jacod-Shiryaev-1987}, Chapter VI, p.315, 3.25 Definition, which we could not obtain.}. Also, by definition, each component of $\BZ_T(\cd)$ is a difference of non-decreasing functions, and
$$\dbE|\BZ_T(t)|\les\(\dbE|\BZ_T(t)|^2\)^{1\over2}=\(\tr\dbE\lan\BZ_T(t),
\BZ_T(t)\ran\)^{1\over2}=\(\dbE[\mathbbm{1}^\top\BX_T(t)]|\)^{1\over2}.$$
Consequently, $\BZ_T(\cd)$ also satisfies the Meyer-Zheng's condition. Moreover,
\bel{X_T-L_T}\BX_T(t)-\BBL_T(t)={1-a_T\over T\m_T}\(\BN_T(tT)-\int_0^{tT}\BBl_T(s)ds\)={1-a_T\over T\m_T}\BM_T(tT),\ee
which is a martingle. By Doob's inequality and the fact \rf{BMBM}, i.e., $\lan\BM_T,\BM_T\ran(t)=\diag\int_0^t\BBl_T(r)dr$,
\bel{X-L}\ba{l}
\ns\ds\dbE\[\sup_{s\in[0.t]}|\BX_T(s)-\BBL_T(s)|^2\]=\({1-a_T\over T\m_T}\)^2\dbE\[\sup_{s\in[0,t]}
|\BM_T(sT)|^2\]\les C\({1-a_T\over T\m_T}\)^2\dbE|\BM_T(tT)|^2\\
\ns\ds\les C\({1-a_T\over T\m_T}\)^2\dbE|\BN_T(tT)|\les{C(1-a_T)\|\BBm_T(\cd)\|_\i t\over T\m_T^2}=C{1-a_T\over\m_T}{\|\BBm_T(\cd)\|_\i t\over T\m_T}\to0.\ea\ee
This implies that the possible limits $\BX(\cd)$ and $\BBL(\cd)$ of $\BX_T(\cd)$ and $\BBL_T(\cd)$ will be the same, and nontrivial (due to \rf{4.5*}) although the factor ${1-a_T\over T\m_T}\to0$.

\ms

Next, let $(\BX(\cd),\BZ(\cd))$ be a limit point of $(\BX_T(\cd),\BZ_T(\cd))$. Since the maximum jump size of $\BX_T(\cd)$ is ${1-a_T\over T\m_T}$ and that of $\BZ_T(\cd)$ is $\sqrt{1-a_T\over T\m_T}$, both are going to zero. Thus, $\BX(\cd)$ and $\BZ(\cd)$ are continuous. Since $\BZ_T(\cd)$ is a martingale, so is $\BZ(\cd)$. Now, for $i=2,3,4$,
\bel{v_iBBL_T(t)}0\les\dbE[v_i^\top\BBL_T(t)]={1-a_T\over T\m_T}\int_0^{tT}\dbE[v_i^\top\BBl_T(r)]dr\to0,\qq T\to\i.\ee
and
\bel{v_iBX_T(t)}\dbE|v_i^\top\BX_T(t)|\les|v_i|\dbE|\BX_T(t)-\BBL_T(t)|+\dbE[v_i^\top\BBL_T(t)]\to0,
\qq T\to\i.\ee
Therefore, for any possible limits $(\BX(\cd),\BBL(\cd))$ of $(\BX_T(\cd),\BBL_T(\cd))$, it holds
\bel{v_iX=0}v_i^\top\BX(t)=v_i^\top\BBL(t)=0,\qq t\in[0,\i),~i=2,3,4.\ee
For $i=1$, we recall that
$$v_1^\top\BBL_T(t)={1-a_T\over T\m_T}\int_0^{tT}v_1^\top\BBl_T(s)ds={1\over T}\int_0^{tT}Y_T({s\over T})ds=\int_0^tY_T(s)ds.$$
Thus, at the limit, one has
$$v_1^\top\BX(t)=v_1^\top\BBL(t)=\int_0^tY(s)ds.$$
Then, recalling \rf{v_iX=0} and \rf{X-L}, we obtain
\bel{L=X}\BBL(t)=\BX(t)={1\over\e_1^\top v_1}\(\int^t_0 Y(s)ds\)\e_1,\qq t\ges0.\ee
Now, note
\bel{ZZ}\lan\BZ_T,\BZ_T\ran(t)={1-a_T\over T\m_T}\lan\BM_T,\BM_T\ran(tT)={1-a_T\over T\m_T}\int_0^{tT}\diag\BBl_T(s)ds=\diag\BBL_T(t).\ee
Thus, at the limit, it holds that
$$\lan\BZ,\BZ\ran(t)=\diag\BBL(t)=\diag\BX(t)={1\over\e_1^\top v_1}\(\int^t_0 Y(s)ds\)I,\qq t\ges0.$$
From the above we derive \rf{Z*} by an argument of \cite{Revus-Yor-2013}, p.203 (see also
\cite{El_Euch-2018}, p.63, \cite{El_Euch-Fukasawa-Rosenbaum-2018} p.273 and \cite{El_Euch-Rosenbaum-2018}, p.26). The rest of the conclusions are clear. \endpf

\subsection{Scaling limit of accumulative HFT volume}

We now observe the following:
$$\sqrt{1-a\over T\m_T}V_T(tT)=\sqrt{1-a_T\over T\m_T}v_4^\top\BN_T(tT)
=\sqrt{1-a_T\over T\m_T}v_4^\top\(\BM_T(tT)+\int^{tT}_0\BBl_T(s)ds\).$$
Furthermore,
$$\ba{ll}
\ns\ds v_4^\top\BBl_T(t)=v_4^\top\(\BBm_T(t)+a_T\int_0^t\BBF_0\f(t-s)d\BN_T(s)\)\\
\ns\ds=v_4^\top\BBm_T(t)+a_T\l_4\int^t_0\f(t-s)d\big(v_4^\top\BN_T(s)\big)\\
\ns\ds=v_4^\top\BBm_T(t)+a_T\l_4\int^t_0\f(t-s)d\big(v_4^\top\BM_T(s)\big)+a_T\l_4\int_0^t
\f(t-s)\big(v_4^\top\BBl_T(s)\big)ds.\ea$$
We have
\bel{vi-CT*}v_4^\top\BBl_T(t)=v_4^\top\BBm_T(t)+\int_0^t\psi_{T,4}(t-s)(v_4^\top
\BBm_T(s))ds+\int_0^t\psi_{T,4}(t-s)d(v_4^\top\BM_T(s)),\ee
Then using Fubini theorem, we get
$$\ba{ll}
\ns\ds\int_0^{tT}v_4^\top\BBl_T(s)ds=\int_0^{tT}v_4^\top\BBm_T(s)ds+\int_0^{tT}\(\int_0^s
\psi_{T,4}(s-r)(v_4^\top\BBm_T(r))dr\)ds\\
\ns\ds\qq\qq\qq\qq\qq+\int_0^{tT}\(\int_0^s\psi_{T,4}(s-r)d\big(v_4^\top \BM_T(r)\big)\)ds\\
\ns\ds\qq\qq\qq\q=\int_0^{tT}v_4^\top\BBm_T(s)ds+\int_0^{tT}\(\int_0^{tT-s}\psi_{T,4}(r)
dr\)(v_4^\top\BBm_T(s))ds\\
\ns\ds\qq\qq\qq\qq\qq+\int_0^{tT}\(\int^{tT-s}_0\psi_{T,4}(r)dr\)d\big(v_4^\top \BM_T(s)\big).\ea$$
Thus,
$$\ba{ll}
\ns\ds\sqrt{1-a_T\over T\m_T}V_T(tT)=\sqrt{1-a_T\over
T\m_T}v_4^\top\BN_T(tT)=\sqrt{1-a_T\over T\m_T}v_4^\top\(\BM_T(tT)+\int^{tT}_0\BBl_T(s)ds\)\\
\ns\ds=\sqrt{1-a_T\over
T\m_T}\[\int^{tT}_0d\big(v_4^\top\BM_T(s)\big)+\int_0^{tT}v_4^\top\BBm_T(s)ds+\int_0^{tT}\(\int_0^{tT-s}\psi_{T,4}(r)
dr\)(v_4^\top\BBm_T(s))ds\\
\ns\ds\qq\qq\qq\qq\qq+\int_0^{tT}\(\int^{tT-s}_0\psi_{T,4}(r)dr\)d\big(v_4^\top \BM_T(s)\big)\]\\
\ns\ds=v_4^\top\BZ_T(t)+\sqrt{1-a_T\over
T\m_T}\[\int_0^{tT}v_4^\top\BBm_T(s)ds+\int_0^{tT}\(\int_0^{tT-s}\psi_{T,4}(r)
dr\)(v_4^\top\BBm_T(s))ds\]\\
\ns\ds\qq-\int_0^{tT} \int^\infty_{tT-s}\psi_{T,4}(r)drd\big(v_4^\top\BZ_T(s)\big)+\int_0^\i\psi_{T,4}(r)dr
\big(v_4^\top\BZ_T(t)\big)\\
\ns\ds=\(1+\int_0^\i\psi_{T,4}(r)dr\)\big(v_4^\top\BZ_T(t)\big)-R_T(t).\ea$$
with
$$\ba{ll}
\ns\ds R_T(t)=\sqrt{1-a_T\over
T\m_T}\[\int_0^{tT}v_4^\top\BBm_T(s)ds+\int_0^{tT}\(\int_0^{tT-s}\psi_{T,4}(r)
dr\)(v_4^\top\BBm_T(s))ds\]\\
\ns\ds\qq\qq\qq-\int_0^{tT}\(\int^\infty_{tT-s}\psi_{T,4}(r)dr\)d\big(v_4^\top\BZ_T(s).\big)\ea$$
Since
$$\int_0^\i\psi_{T,4}(r)dr=\int_0^\i\sum_{k=1}^\i(a_T\l_4)^k\f^{*k}(r)dr=\sum_{k=1}^\i\big(
a_T\l_4\|\f(\cd)\|_1\big)^k={a_T\l_4\|\f(\cd)\|_1\over1-a_T\l_4\|\f)(\cd)]\|_1}.$$
Therefore, we have
$$\sqrt{1-a_T\over T\m_T}V_T(tT)={1\over1-a_T\l_4\|\f(\cd)\|_1}\big(v_4^\top\BZ_T(t)\big)-R_T(t).$$
Now, we estimate $R_T(t)$.
$$\ba{ll}
\ns\ds\dbE|R_T(t)|^2\les K\[{1-a_T\over T\m_T}\(\int_0^{tT}|v_4^\top\BBm_T(s)|ds\)^2+\dbE\Big|\int_0^{tT}
\(\int_{tT-s}^\i\psi_{T,4}(r)dr\)d\big(v_4^\top\BZ_T(s)\big)\Big|^2\]\\
\ns\ds\les K\Big\{{(1-a_T)\m_T\over
T}\[\int_0^{tT}\Big|v_4^\top\({\BBm_T(s)\over\m_T}-\mathbbm{1}\)\Big|ds\]^2
+\int_0^{tT}\Big|\(\int_{tT-s}^\i\psi_{T,4}(r)dr\)\Big|^2v_4^\top\big[\dbE\lan \BZ_T,\BZ_T\ran(s)\big]v_4ds\Big\}\\
\ns\ds\les K\Big\{T^{-\a}(\bar a+o(1))T^{\a-1}(\bar\m+o(1)) t\[\int_0^{tT}\3n\({|\BBm_T(s)|\over\m_T}-\mathbbm{1}\)^2ds\]\\
\ns\ds\qq\qq+{1-a_T\over T\m_T}\int_0^{tT}\Big|\(\int_{tT-s}^\i\3n\psi_{T,4}(r)dr\)\Big|^2|\dbE\BBl_T(s)|ds\Big\}\equiv I_1+I_2.\ea$$
By our assumption, for any $\e>0$,  there exists a $T_0>0$ such that
$$\Big|{\BBm_T(s)\over\m_T}-\mathbbm{1}\Big|<\e,\qq s\ges T_0.$$
Consequently,
$$I_1\les{KT_0\over T}+{K\e^2(tT-T_0)\over T}.$$
This proves that $I_1$ goes zero as $T\to\i$. Also, by the dominated convergence theorem,
$$I_2={1-a_T\over \m_T}\int_0^t\Big|\(\int_{(t-s)T}^\i\3n\psi_{T,4}(r)dr\)\Big|^2|\dbE\BBl_T(sT)|ds\les C\int_0^t\Big|\(\int_{(t-s)T}^\i\3n\psi_{T,4}(r)dr\)\Big|^2ds\to0.$$\
Hence,
$$\ba{ll}
\ns\ds\lim_{T\to\i}\sqrt{1-a_T\over T\m_T}V_T(tT)=\lim_{T\to\i}\(1+\int_0^\i\psi_{T,4}(r)dr\)\big(v_4^\top\BZ_T(t)\big)\\
\ns\ds=\lim_{T\to\i}\(1+\int_0^\i\sum_{k=1}^\i(a_T\l_4)^k\f^{*k}(r)dr\){2\over\e_1^\top v_1}\int_0^t\sqrt{Y(s)}d\big(v_4^\top\BB(s)\big)\\
\ns\ds={4\over\sqrt{\e_1^\top v_1}\big(1-\l_4\|\f(\cd)\|_1\big)}\int_0^t\sqrt{Y(s)}dW(s),\ea$$
where
$$W(s)={1\over2}v_4^\top\BB(s),\qq s\ges0,$$
is a one-dimensional standard Brownian motion. The above leads to the following corollary.

\ms

\bc{cor3.1} \sl Let
$$h(T)=\sqrt{1-a_T\over T}.$$
Then
$$\lim_{T\to\i}h(T)V_T(tT)=c\int_0^t\sqrt{Y(s)}dW(s)\equiv f(t),\q t\ges0,$$
for some function positive $c(\cd)$, and one-dimensional standard Brownian motion $W(\cd)$.
\ec

\section{Dynamics of the LOB with Hawkess Processes}

Combining the esults in the above section, we obtain the following system for the LOB volume density:
\bel{SPDE-system}\left\{\2n\ba{ll}
\ns\ds du(t,x)=\[\eta u_{xx}(t,x)-\b\sgn(x)[u_x(t,x)]^--\z u(t,x)+J(x,u(t,x))+G\(x,\int^{\iota}_{-\iota}u(t,y)dy\)\]dt\\
\ns\ds\qq\qq\qq+cu(t,x)\sqrt{Y(t)}dW(t),\qq t\ges0,~x\in(-L,L),\\
\ns\ds u(0,x)=u_0(x),\qq x\in[-L,L],\\
\ns\ds u(t,\pm L)=0,\qq t\ges0,\\
\ns\ds Y(t)={\bar\n\over\G(\a)}\int_0^t(t-s)^{\a-1}\big(v_1^\top\mathbbm{1}-Y(s)\big)ds
+{\bar\k\bar\n\over\G(\a)}\int_0^t(t-s)^{\a-1}\sqrt{Y(s)}dB_1(s),\q t\ges0.\ea\right.\ee
Also, we recall that
\bel{J*}J(x,u(t,x))=-\sgn(x)\big[j(|x|)\big(u(t,|x|)-u_0\big)^++j(-|x|)\big(u(t,-|x|)+u_0
\big)^-\big],\ee
with $j(x)$ being positive and increasing for $x<0$, decreasing for $x>0$. Also, $G(x,\ell)$ is strictly decreasing in $\ell$, with $G(x,0)=0$. Now, we let $H=L^2(-L,L)$ and $A:\sD(A)\equiv W^{2,2}(-L,L)\cap W^{1,2}_0(-L,L)\subset H\to H$ be a linear densely defined operator on $H$ as
$$Au=-\eta u_{xx}+\z u,$$
with $\eta,\z$ positive constants. With the Direchlet boundary condition, it is know that $A$ is self-adjoint, positive-definite, and $-A$ generates a $C_0$-semigroup $e^{-At}$ which is analytic (see \cite{Pazy-1983}). Further, for any $\g\in\dbR$, $A^\g$ can be well-defined, satisfying
$$\|Ae^{-At}\|\les{C\over t^\g},\qq t>0.$$
We denote $H^\g=\sD(A^\g)$ with the norm $|\xi|_{H^\g}=|A^\g\xi|_H$ for all $\xi\in H^\g$. See \cite{Engel-Nagel-2000}, Chapter 2, Section 5, for details. Also, we denote
$$F(u(t))=-\b\sgn(\cd)[u_x(t,\cd)]^-+J(\cd\,,u(t))+G\(\cd\,,\int^{\iota}_{-\iota}u(t,y)dy\).$$
Then the first equation in \rf{SPDE-system} can be written as
\bel{Evolution}du(t)=\big[-Au(t)+F(u(t))\big]dt+cu(t)\sqrt Y(t)dW(t),\qq t\ges0.\ee
We introduce the following notion. A process $u(\cd)$ is called a mild solution to the above, if the following holds
\bel{u(t)}u(t)=e^{-At}u_0+\int_0^te^{-A(t-s)}\big[\wt F(u(t))+F(u(s))\big]ds+\int_0^te^{-A(t-s)}cu(s)\sqrt{Y(s)}dW(s),\qq t\ges0.\ee

\ms

It is known that the stochastic Volterra integral equation in \rf{SPDE-system} admits a unique solution $Y(\cd)$. Next, is easy to see that
$$0\les\dbE Y(t)\les{\bar\n v_1^\top\mathbbm{1}\over\a\G(\a)}t^\a\equiv Ct^\a,\qq t\ges0.$$
We now define
$$\t_k=\inf\{t>0\bigm|Y(t)>k\},\qq k>0.$$
Then, $\t_k$ is a stopping time and for each $K>0$,
$$\dbP(\t_k>K)=\int_{(\t_k>K)}d\dbP\les\int_{(\t_k>K)}{Y(t)\over K}d\dbP\les{Ct^\a\over K}.$$
Hence, for given finite time horizon $T>0$, one has
$$\lim_{k\to\i}\t_k\land T=T,$$
and
$$0\les Y(t)\les k,\qq t\in[0,\t_k\land T].$$
Therefore, to establish the well-posedness of \rf{SPDE-system}, it suffices to show that \rf{Evolution} is well-posed (for any initial state $u_0$) over $[0,\t_k]$.

\ms

Now, the SPDE in \rf{SPDE-system}, or \rf{Evolution}, on $[0,\t_k]$ is exactly covered by the result of \cite{Da_Prato-Jentzen-Rockner-2019}, p.3771. Thus, we have the following result.

\bt{} \sl Under the above framework, SPDE \rf{Evolution} on $[0,\t_k]$ admits a unique solution $u(\cd)$, with
\bel{}\sup_{t\in[0,\t_k]}\dbE\[|u(t)|_{H^{1\over2}}^2\]<\i.\ee

\et

\section{Price Dynamics}

The bid and ask price dynamics are determined by the LOB dynamics.  When the ask (bid) queue is depleted, the price moves up (down) to the next level of the order book.  We assume that the order book contains no gaps so that the price increments are equal to one tick, which is $\delta$ as defined in Section 3.1.  When the bid queue is depleted, the price decreases by one tick. When the ask queue is depleted, the price increases by one tick.  On the other hand, if the queue sizes increase rapidly in a short period of time, it means there are excessive amount of limit orders, which will likely be transferred to market orders and be executed towards the opposite direction.  When the ask queue size increases $n$ times, the price will move down $n$ ticks.  When the bid queue size increases $n$ times, the price will move up $n$ ticks.

\ms

We use a simple example to illustrate how the LOB dynamics determine the bid and ask prices.  Suppose in the LOB below, there is a bid order of 10,000 shares, then the first 2 queues on the ask side will be depleted, and the ask price will moving up 2 ticks, rising from \$100.01 to \$100.03. All 3 LOB activities affect the ask and bid queues.  Submission of limit orders increase the queues, while cancellation of limit orders as well as market orders from the opposite side decrease the queues. Therefore, the price changes are determined by the volume changes, and we model the volume by the order book depth \cite{Cont-Mueller-2021}.

\ms

Let $D^a(t)$ ($D^b(t)$) be the volume of limit ask (bid) orders at the top of the LOB at time $t$.  The order book depth can be expressed as
$$D^a(t)=\int_0^\iota u(t,x)dx,\qq D^b(t)=\int_{-\iota}^0u(t,x)dx$$
Let the change of the order book depth in the time interval $[t, t+dt]$ be $dD^a(t)$ and $dD^b(t)$.  Note that since $u(t,x)>0$ on the ask side and $u(t,x)<0$ on the bid side, $D^a(t)>0$ and $D^b(t)<0$. We now discuss the situation in a little details below.

\ms

$\bullet$ When $dD^a(t)<0$, the ask queue decreases and the ask price increases by $-C_a{dD^a(t)\over D^a(t)}$ ticks, with some constant $C_a>0$. When $dD^a(t)>0$, the ask price decreases by $C_a{dD^a(t)\over D^a(t)}$ ticks. Therefore, the price impact from the ask queue is $-C_a{dD^a(t)\over D^a(t)}$.

\ms

$\bullet$ When $dD^b(t)>0$, $D^b(t)$ increases, but since $D^b(t)<0$, this means that the bid queue decreases, and the bid price decreases by $-C_b{dD^b(t)\over D^b(t)}$ ticks, for some constant $C_b>0$. When $dD^b(t)<0$, the bid queue increases, and the bid price increases by $C_b{dD^b(t)\over D^b(t)}$ ticks. From the above, we see that the price impact from the bid queue is $C_b{dD^b(t)\over D^b(t)}$.

\ms

In summary, the ask and bid price changes will be:
$$ds^a(t)=-\d C_a{dD^a(t)\over D^a(t)},\qq ds^b(t)=\d C_b{dD^b(t)\over D^b(t)}$$
and the price change will be
$$dS(t)={1\over2}(ds^a(t)+ds^b(t))={\d\over2}\(C_b{dD^b(t)\over D^b(t)}-C_a{dD^a(t)\over D^a(t)}\).$$
We do not assume any condition on $C_a$ and $C_b$. Let us now find the dynamics of $D^a(t)$ and $D^b(t)$. Clearly,
$$\ba{ll}
\ns\ds dD^a(t)=d\int_0^\iota u(t,x)dx=\int_0^\iota du(t,x)dx\\
\ns\ds=\int_0^\iota\Big\{\[\eta u_{xx}(t,x)-\z u(t,x)+F(x,u(t,\cd)\]dt+\int_0^\iota c u(t,x)\sqrt{Y(t)}dW(t)\Big\}dx\\
\ns\ds=\int_0^\iota\Big\{\Big[\eta u_{xx}(t,x)-\zeta_a u(t,x)+F(x,u(t,\cd))\]dt\Big\}dx+ c\(\int_0^\iota u(t,x)dx\)\sqrt{Y(t)}dW(t)\\
\ns\ds=\int_0^\iota\Big\{\[\eta u_{xx}(t,x)-\z u(t,x)+F(x,u(t,\cd))\Big]dt\Big\}dx+  cD^a(t)\sqrt{Y(t)}dW(t).\ea$$
Similarly,
$$dD^b(t)=\int_{-\iota}^0\Big\{\Big[\eta u_{xx}(t,x)-\z u(t,x)+F(x,u(t,\cd))\Big]dt\Big\}dx+cD^b(t)\sqrt{Y(t)}dW(t).$$
Therefore, we have the price dynamics model as
\bel{price}\left\{\2n\ba{ll}
\ds dS(t)={\d\over2}[\n_b(t)-\n_a(t)]dt+{\d c\over2}(C^b-C^a)\sqrt{Y(t)}dW(t),\qq S(0)=S_0>0,\\
\ns\ds Y(t)={\bar\n\over\G(\a)}\int_0^t(t-s)^{\a-1}\big(v_1^\top\mathbbm{1}-Y(s)\big)ds
+{\bar\k\bar\n\over\G(\a)}\int_0^t(t-s)^{\a-1}\sqrt{Y(s)}dB_1(s),\q t\ges0.\ea\right.\ee
where
$$\ba{ll}
\ns\ds\n_a(t)={C_a\over D^a(t)}\int_0^\iota\[\eta u_{xx}(t,x)-\z u(t,x)+F(x,u(t,\cd)) \]dx,\\
\ns\ds\n_b(t)={C_b\over D^b(t)}\int_{-\iota}^0\[\eta u_{xx}(t,x)-\z u(t,x)+F(x,u(t,\cd))\]dx.\ea$$
Since $\n_a(\cd)$ and $\n_b(\cd)$ are depending on $u(\cd\,,\cd)$, the above system is not closed. Therefore, essentially, we need to solve \rf{SPDE-system} together with \rf{price} to get the price process $S(\cd)$.

\section{Numerical Tests}
In this section, we look into our model numerically based on existing literature, then point out potential advantages of our model, comparing it to the C-M model and some other existing models, from the perspective of real-market data.

We divide our model into two parts (without stating the initial and boundary conditions) as below: the stochastic partial differential equation (SPDE)
\begin{align}
d u(t, x) & =\left[\eta u_{x x}(t, x)-\beta \operatorname{sgn}(x)\left[u_x(t, x)\right]^{-}-\zeta u(t, x)+J(x, u(t, x))+G(x, \ell(t))\right] d t \label{eq20241113_1}\\
& +c u(t, x) \sqrt{Y(t)} d W(t), \quad t \ges 0, x \in(-L, L),\notag
\end{align}
and the singular stochastic integral equation (SSIE)
\begin{align}\label{eq20241113_2}
Y(t) =\frac{\bar{\nu}}{\Gamma(\alpha)} \int_0^t(t-s)^{\alpha-1}\left(v_1^{\top} \mathbbm{1}-Y(s)\right) d s
+\frac{\bar{\kappa} \bar{\nu}}{\Gamma(\alpha)} \int_0^t(t-s)^{\alpha-1} \sqrt{Y(s)} d B_1(s).
\end{align}

The numerical challenges of solving the SPDE \eqref{eq20241113_1} and SSIE \eqref{eq20241113_2} are in the following.
\begin{enumerate}
\item For the SPDE \eqref{eq20241113_1}, the noise is smooth in space and white in time, but multiplicative, and the diffusion term depends on $Y(t)$ that is not smooth in time. In the existing literature for stochastic ODEs (and some stochastic PDEs), stability and convergence of the numerical solutions are established when the diffusion term is globally Lipschitz continuous and satisfies the linear growth condition. We refer the readers to references \cite{kloeden1991numerical}, \cite{higham2002strong}, \cite{mao2007stochastic} for the case of stochastic ODEs, and references \cite{gyongy2005discretization}, \cite{hutzenthaler2010strong}, \cite{gyongy2016convergence}, \cite{feng2017finite}, \cite{majee2018optimal}, \cite{beccari2019strong}, \cite{jentzen2015strong}) for the case of stochastic PDEs. Here the diffusion term fails the assumptions such that stability and convergence are unknown.
\item The drift term is nonlinear in the SPDE \eqref{eq20241113_1}. As far as we know in existing literature, to establish the stability and convergence of the numerical solutions, the drift term should be globally Lipschitz continuous or one-sided Lipschitz continuous (nonlinear but behaves polynomially together with some other growth conditions). We refer the readers to references \cite{kloeden1991numerical}, \cite{gyongy2005discretization}, \cite{gyongy2016convergence} for global Lipschitz case and \cite{higham2002strong}, \cite{mao2007stochastic}, \cite{hutzenthaler2010strong}, \cite{feng2017finite}, \cite{majee2018optimal}, \cite{beccari2019strong}, \cite{jentzen2015strong} for one-sided Lipschitz case. The drift term fails these assumptions.
\item For the SPDE \eqref{eq20241113_1}, when the changes of limit order volume density are small and/or the amount of volume density that got canceled at a distance $x$ from the mid-price is large, i.e., $\beta/\eta$ is large, the problem could be convection dominated. In this case, standard discretization methods cannot be used, and special methods should be designed for this case to avoid spurious oscillations in the numerical solutions. This type of problem was investigated by many references, so we do not list all the references here. We just refer the readers to monographs \cite{morton1996numerical}, \cite{roos2008robust} for a detailed description of the numerical techniques. 
\item The SSIE \eqref{eq20241113_2} is a singular integral equation of the second kind. Some existing numerical methods could be employed, i.e., the Galerkin-Petrov method, the Collocation method, the discrete Galerkin method, etc. The discretized formulation of this SSIE will result in an ill-conditioned linear system such that regularization techniques or preconditioning methods are required to improve the condition number of our discretized formulation. The singularity of the SSIE \eqref{eq20241113_2} in the kernel has different intensity based on the values of $\alpha$, and then appropriate strategies should be chosen to improve the accuracy, i.e., regularization, singularity subtraction, Cauchy principal value, appropriate quadrature rules, etc.
\item The regularity of the solutions of the SPDE \eqref{eq20241113_1} and the SSIE \eqref{eq20241113_2} is not strong enough to get any theoretical numerical results. For example, the solution of the SPDE \eqref{eq20241113_1} is $H^{\frac12}$ in space, which is too weak to establish the stability results, high moment results, or the error estimates of the numerical methods, including finite element methods, finite difference methods, discontinuous Galerkin methods, and so on.
\item The SSIE \eqref{eq20241113_2} involves the stochasticity. The computational complexity is higher than the deterministic case. We skip this efficiency discussion for stochastic computation which has been a hot topic for a few years. Not only the efficiency is complex, but also the accuracy is challenging and unknown. There were some references investigating the numerics for the singular integral equations (SIE), but very few results discuss the numerics for the singular stochastic integral equation (SSIE). Based on existing literature \cite{richard2021discrete}, \cite{li2022numerical} (the assumptions on the integrations of the kernels are made in \cite{richard2021discrete}), we summarize the minimal assumptions on the SSIE \eqref{eq20241113_2} to guarantee the convergence of the numerical methods (possibly no order depending on the intensity of the singularity):
\begin{enumerate}
\item Drift term is Lipschitz continuous and satisfies linear growth condition
\item The derivative of the drift term is Lipschitz continuous
\item Diffusion term is Lipschitz continuous and satisfies linear growth condition
\item p-moment of the strong solution is bounded
\end{enumerate}
Our SSIE \eqref{eq20241113_2} fails the assumptions, so the numerical convergence of our SSIE \eqref{eq20241113_2} is another open problem.
\end{enumerate}

Due to the above numerical challenges and some of them are open problems, we leave the numerics for our model as open problems to interesting readers in the area of scientific computing and stochastic computation. Instead, we do some tests showing the potential advantages of our proposed model, comparing to the C-M model in reference \cite{Cont-Mueller-2021} and other similar models.

\medskip
{\bf Test 1.} Depth of the Order Book.

\begin{figure}[H] 
\centering
\includegraphics[height=4.5in,width=5.5in]{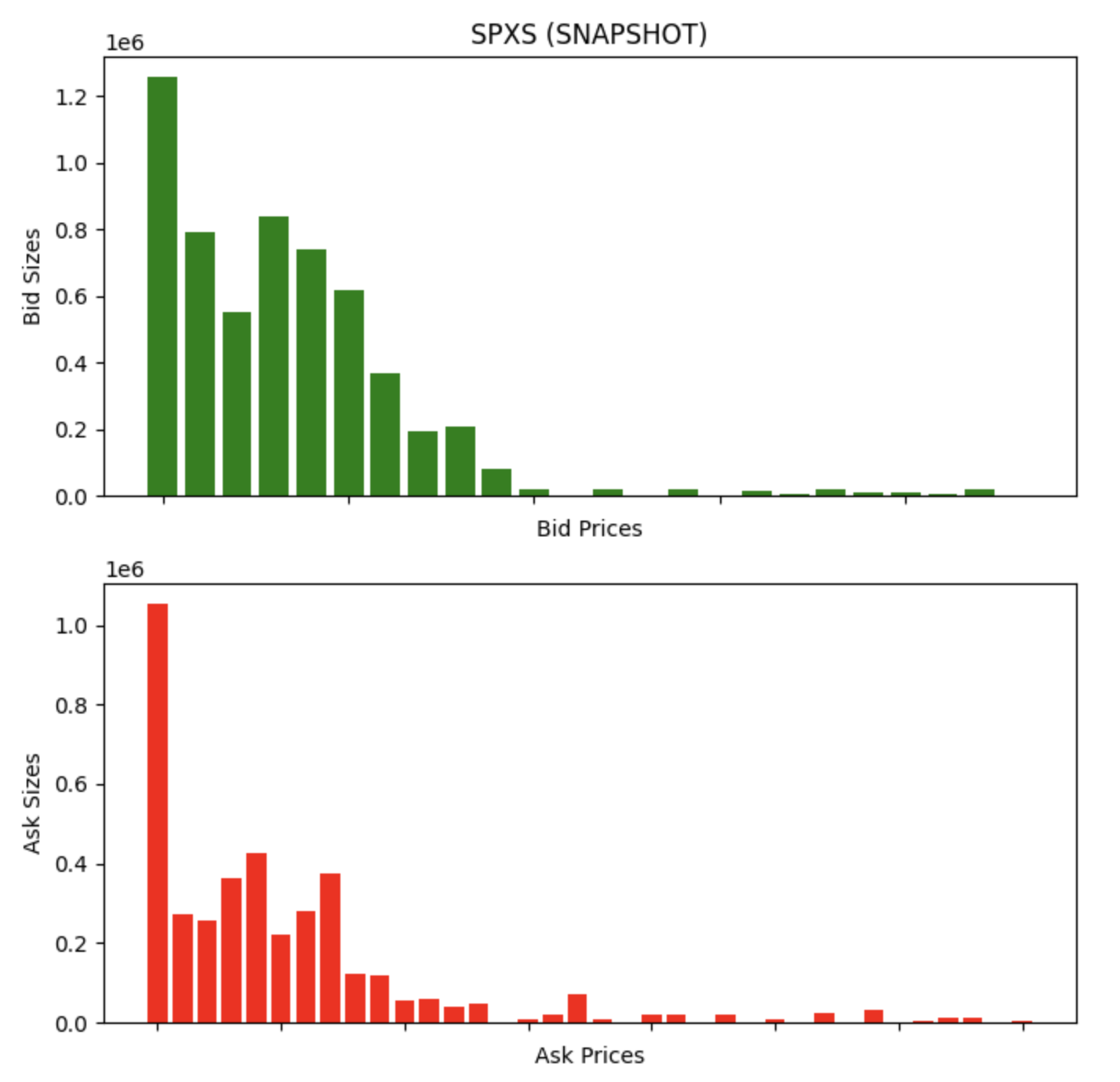}
\caption{A snapshot of the model bid sizes and model ask sizes for the ticker symbol `SPXS'.}
\label{fig1_1}
\end{figure}

\begin{figure}[H] 
\centering
\includegraphics[height=3.8in,width=5.5in]{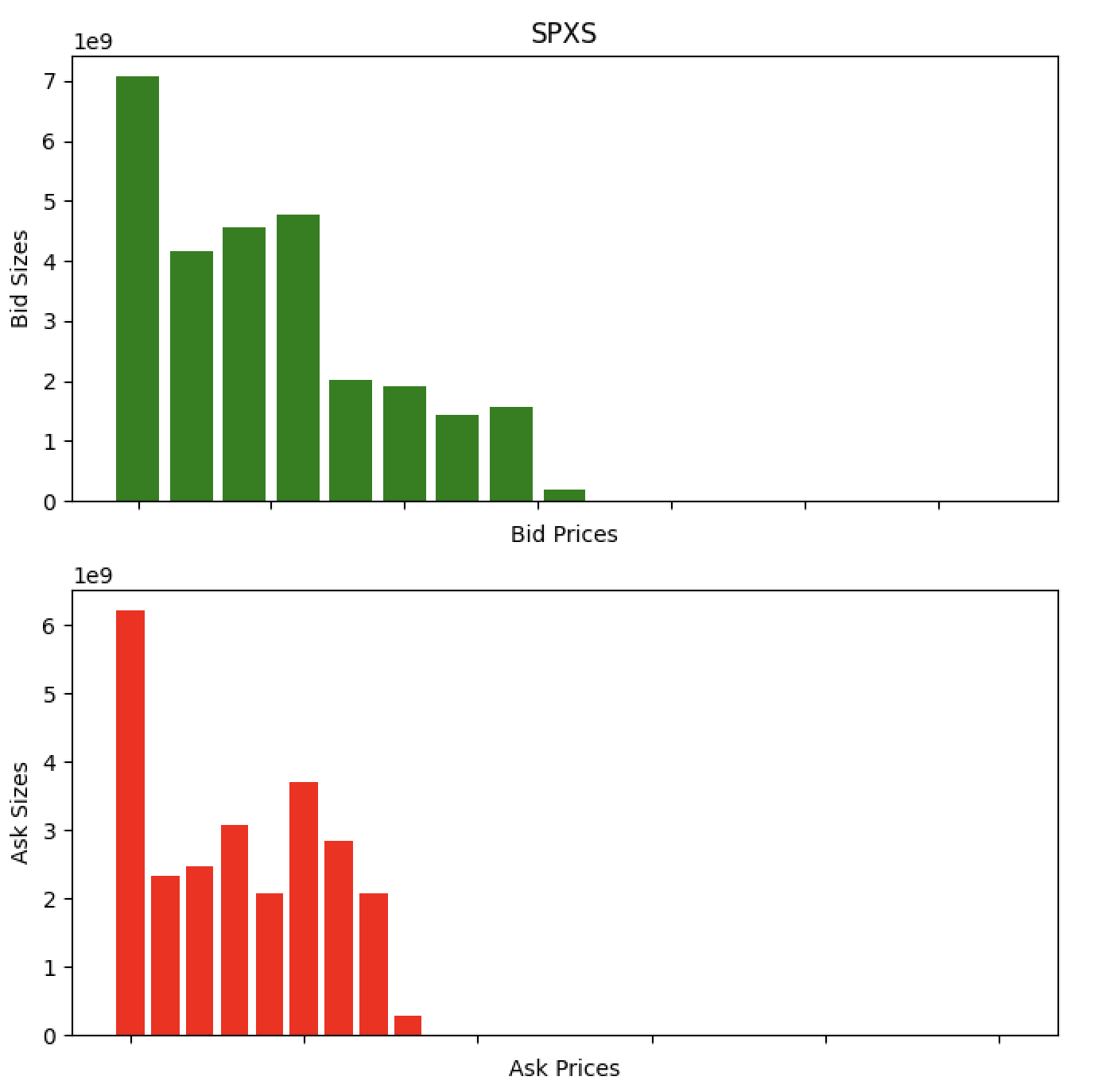}
\caption{The sum of the model bid and model ask over several minutes when stock prices of `SPXS' keep the same.}
\label{fig1_2}
\end{figure}

\begin{figure}[H] 
\centering
\includegraphics[height=3.8in,width=5.5in]{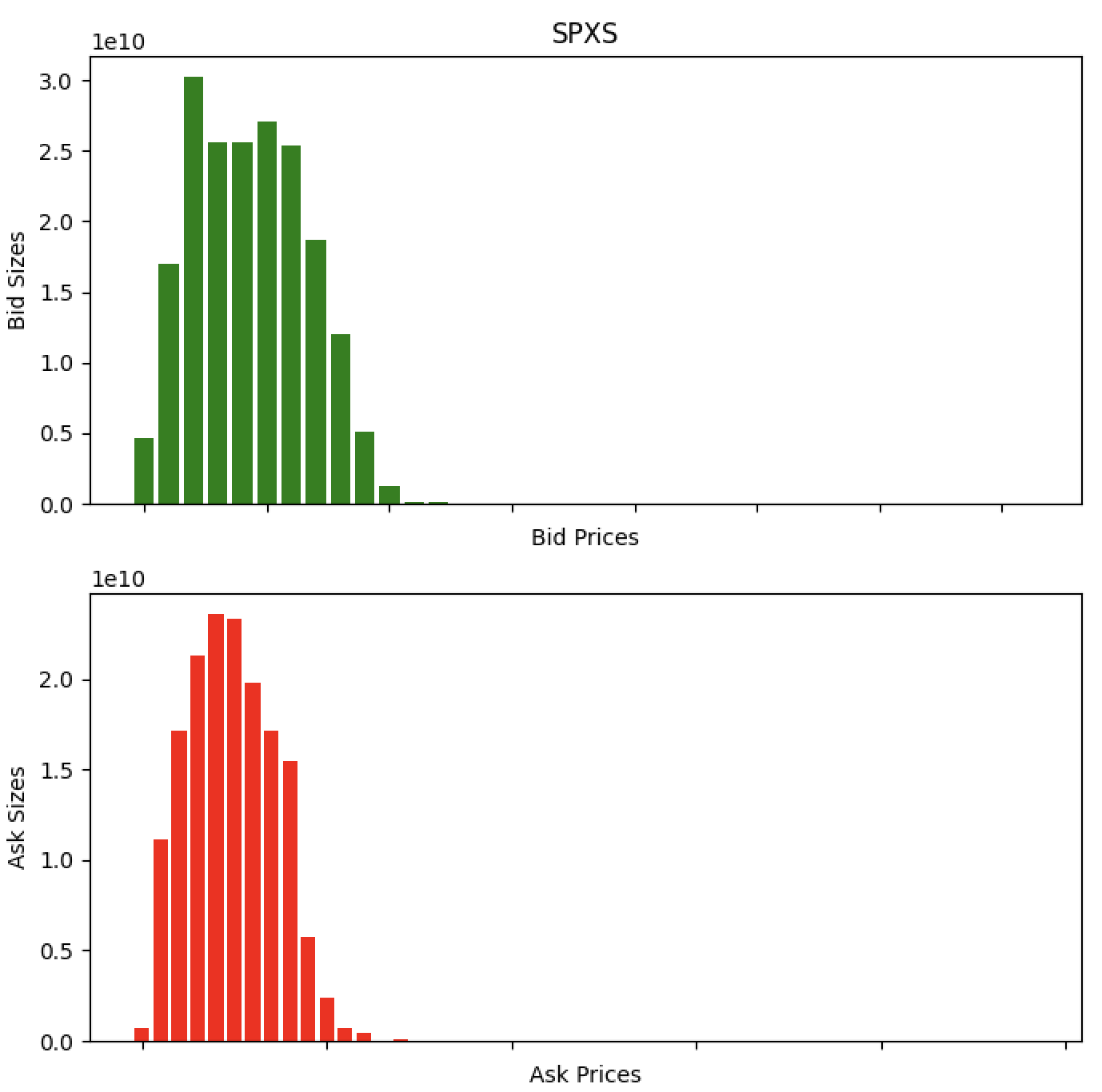}
\caption{The sum of the model bid and model ask over longer time when stock prices of `SPXS' change.}
\label{fig1_3}
\end{figure}

In this test, the market depth of the order book is considered. The model bid and model ask in the real market order book could possibly generate several different patterns. Our model has nonlinear structures in the convection term and other nonlinear terms, i.e., the truncation/sign functions indicate the bid orders and ask orders. The nonlinear model matches with the real market data and could potentially capture different patterns in different scenarios. The following are different patterns of data-bid and data-ask based on the real market data. Various patterns even random ones could be obtained based on different ticker symbols or the same ticker symbol during different time periods. Therefore, we consider the snapshot with large model bid sizes and ask sizes to avoid random patterns. In Figures \ref{fig1_1}-\ref{fig1_3}, bid prices decrease from the left to the right, and the ask prices increase from the left to the right. Figure \ref{fig1_1} is a snapshot of the model bid/ask sizes with respect to the bid/ask prices for the ticker symbol `SPXS'. Figure \ref{fig1_2} shows a pattern over several minutes. The model bid and model ask keep the same such that the model bid and model ask closest to the stock price dominate. Figure \ref{fig1_3} shows another pattern over a longer time. The model bid and model ask move a lot such that the graphs look like log-normal distribution. Note that the bid sizes and ask sizes in the tails are not zeros since they are just relatively small comparing to the peak sizes.

\medskip
{\bf Test 2. The Volatility.}

The rough volatility originates from the empirical observation that log-volatility behaves like a fractional Brownian motion, and it is a very important concept in financial mathematics to model the volatility of financial assets, especially for short-term fluctuations. Its advantages include better capturing complex realistic market dynamics, better pricing and hedging derivatives, robustness to high-frequency noise, power for predictivity, and so on.

\begin{figure}[H]
\centering
\includegraphics[height=3.2in,width=5.0in]{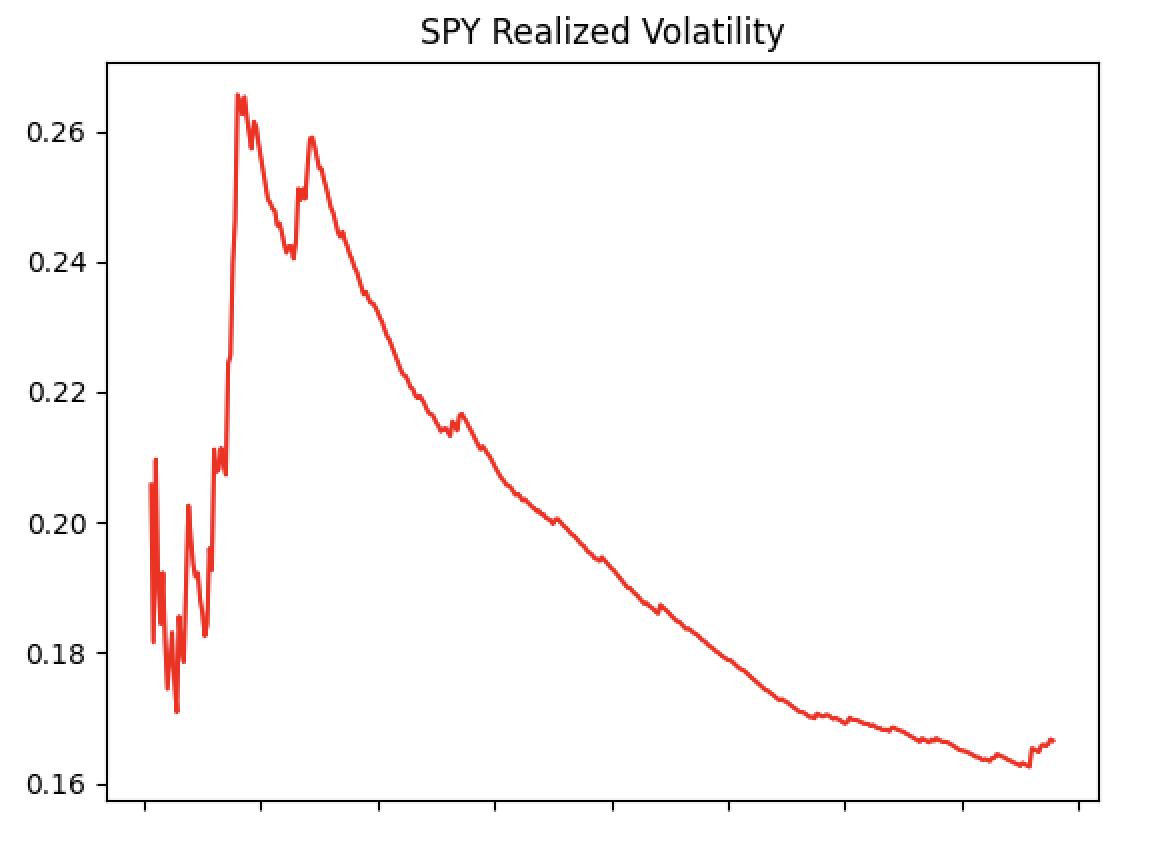}
\caption{Minute-level Realized Volatility of SPY from 2024-11-13 08 AM. to 2024-11-13 10 AM.}
\label{fig2_1}
\end{figure}

\begin{figure}[H]
\centering
\includegraphics[height=3.2in,width=5.0in]{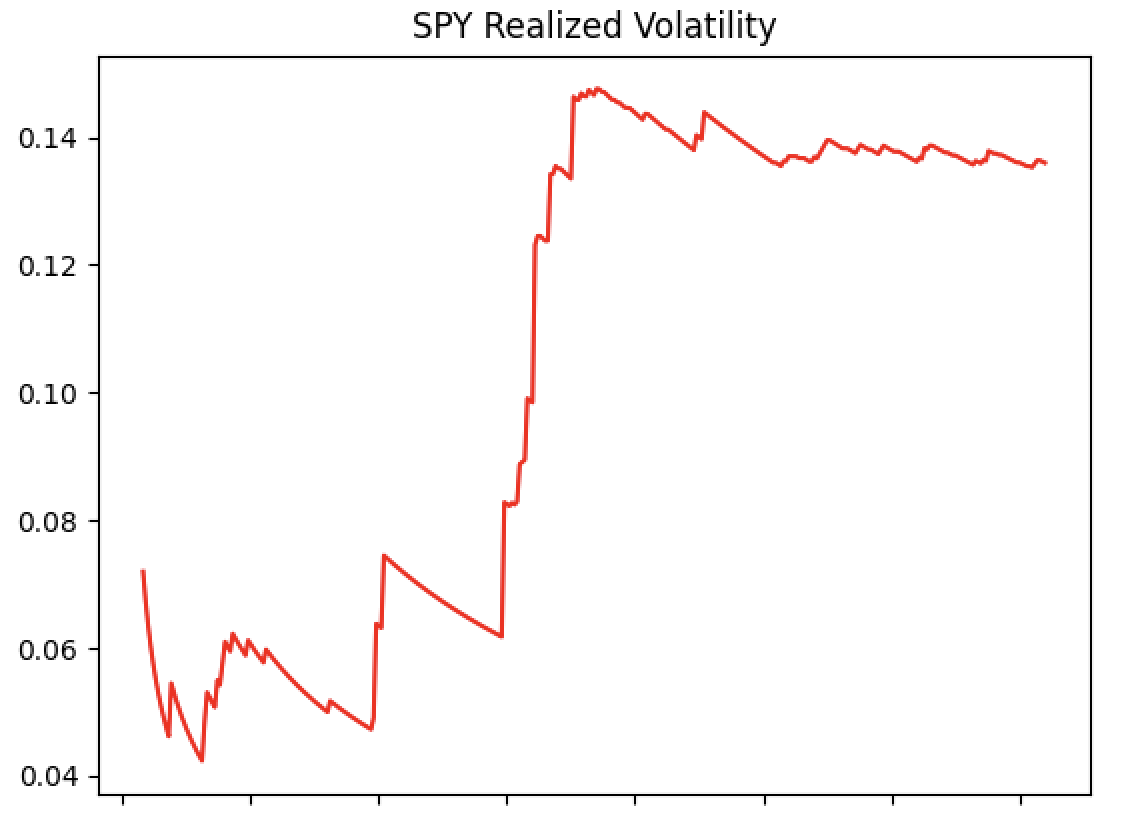}
\caption{Microsecond-level Realized Volatility of SPY around 2024-11-26 14:00:00.}
\label{fig2_2}
\end{figure}

Here we do some tests showing that realized volatility often exhibits characteristics of ``roughness" in real market data, which illustrates from another perspective that our model could perform better than the constant volatility in \cite{Cont-Mueller-2021}. For example, Figure \ref{fig2_1} uses the close prices to plot the realized volatility of SPY. During the 2-hour time period, there were some events in the beginning such that the stock prices were more volatile and hence the realized volatility looked very rough. Figure \ref{fig2_2} shows the realized volatility of SPY based on the HFT data. 
We can see that the volatility increases much immediately after a time point, and before that, the volatility is relatively small but oscillates.

Besides, the ``roughness" of the realized volatility can also be observed based on the real-time data in some other aspects. For example, the autocorrelations of increments in realized volatility show a power-law decay, the empirical studies carried out by some existing references, and so on.

\section{Conclusion}

In this paper, we have derived an HFT model in which the market orders are allowed and the scaling limit of the multi-dimensional (self-exciting) nearly unstable Hawkes process with power tails has been used to describe the HFT order.

\ms

Based on the order book dynamics, we also created a middle price dynamics model in the same market. We can see that among all the parameters, $\alpha$, the parameter that measures how frequently the metaorder splitting strategy is used, has the most significant impact. It turns out that the more frequently the strategy is used, the larger volatility there will be in the price change.

\appendix
\section{Technical Appendix}

\ms

In this appendix, we calculate the eigenvalues of $\BBF_0$ (and $\BBF_0^\top$), where
$$\BBF_0=\begin{pmatrix}
    1 & 0 & \b_2 & (\b_1+\b_2+\b_2\b_3-1)\\
    0 & 1 &(\b_1+\b_2+\b_2\b_3-1)&\b_2\\
    \b_2&\b_2\b_3 & (\b_1+\b_2) & 0 \\
     \b_2\b_3&\b_2& 0 & (\b_1+\b_2)
    \end{pmatrix}$$
Let the eigenvalue be $\l$.
$$\ba{ll}
\ns\ds\det(\BBF_0-\l I)=\begin{vmatrix}
    1-\l & 0 & \b_2 & \b_1+\b_2+\b_2\b_3-1 \\
    0 & 1-\l & \b_1+\b_2+\b_2\b_3-1 & \b_2 \\
    \b_2 & \b_2\b_3 & \b_1+\b_2-\l & 0\\
    \b_2\b_3 & \b_2 & 0 & \b_1+\b_2-\l\end{vmatrix}\\
\ns\ds=\l^4-(2\b_1+2\b_2+2)\l^3+(\b_1^2+4\b_1-\b_2^2+2\b_1\b_2+4\b_2-2\b_2^2\b_3^2
-2\b_2^2\b_3-2\b_1\b_2\b_3+2\b_2\b_3+1)\l^2\\
\ns\ds-(2\b_1^2+2\b_1-2\b_2^3-2\b_1\b_2^2+4\b_1\b_2+2\b_2-2\b_2^3\b_3^2-2\b_1\b_2^2\b_3^2 -2\b_2^2\b_3^2-2\b_2^3\b_3-4\b_1\b_2^2\b_3\\
\ns\ds-2\b_1^2\b_2\b_3+2\b_2\b_3)\l\\
\ns\ds+(\b_1^2-2\b_1\b_2^3-\b_1^2\b_2^2+2\b_1\b_2+\b_2^4\b_3^4+2\b_2^4\b_3^3+ 2\b_1\b_2^3\b_3^3-2\b_2^3\b_3^3-\b_2^4\b_3^2+2\b_1\b_2^3\b_3^2-4\b_2^3\b_3^2\\
\ns\ds+\b_1^2\b_2^2\b_3^2-4\b_1\b_2^2\b_3^2+\b_2^2\b_3^2-2\b_2^4\b_3- 2\b_1\b_2^3\b_3 -4\b_1\b_2^2\b_3+2\b_2^2\b_3-2\b_1^2\b_2\b_3+2\b_1\b_2\b_3)\\
\ns\ds=\(\l-(\b_2-\b_2\b_3+1)\)\(\l^3-(2\b_1+\b_2+\b_2\b_3+1)\l^2+ (\b_1^2+2\b_1-2\b_2^2+2\b_2-\b_2^2\b_3^2\\
\ns\ds-2\b_2^2\b_3+2\b_2\b_3)\l-(\b_1^2-2\b_1\b_2^2-\b_1^2\b_2+2\b_1\b_2- \b_2^3\b_3^3-3\b_2^3\b_3^2-2\b_1\b_2^2\b_3^2+\b_2^2\b_3^2\\
\ns\ds-2\b_2^3\b_3-4\b_1\b_2^2\b_3+2\b_2^2\b_3-\b_1^2\b_2\b_3+2\b_1\b_2\b_3)\)\\
\ns\ds=\(\l-(\b_2-\b_2\b_3+1)\)\(\l+(\b_2+\b_2\b_3-1)\)\(\l-(\b_1+\b_2\b_3)\)\( \l-(\b_1+2\b_2+\b_2\b_3)\)=0.\ea$$
Therefore, we get the eigenvalues
$$\l_1=\b_1+\b_2\b_3+2\b_2,\q\l_2=-\b_2\b_3+\b_2+1,\q\l_3=\b_1+\b_2\b_3,\q\l_4
=-\b_2\b_3-\b_2+1.$$

\ms

\bf Acknowledgement. \rm The authors would like to thank the anonymous referees for their suggestive comments which lead to this improved version.

\end{document}